\documentclass[amsmath,amssymb,twocolumn,nofootinbib]{revtex4-1}
\usepackage{color}
\usepackage[dvipsnames]{xcolor}
\usepackage{graphicx,epsfig,hyperref,latexsym}
\usepackage{amsmath}
\usepackage{mathrsfs}
\usepackage{braket}
\usepackage[normalem]{ulem}
\usepackage{mathtools}
\usepackage{ulem}
\usepackage{orcidlink}

\usepackage[T1]{fontenc}
\usepackage[utf8]{inputenc}

\newcommand\eq[1]{\begin{equation}\begin{aligned}#1\end{aligned}\end{equation}}
\newcommand\eref[1]{(\ref{eq:#1})}
\newcommand\del[2]{\frac{\partial#1}{\partial#2}}
\newcommand\I{\mathrm{i}}
\newcommand\D{\mathrm{d}}
\newcommand\Ob{\mathcal{O}}
\newcommand\E{\mathcal{E}}
\newcommand\e{\mathrm{e}}
\newcommand\V{\mathcal{V}}

\begin{document}

\title{Phase time and Ehrenfest's theorem\\ in relativistic quantum mechanics and quantum gravity}

\author{Leonardo Chataignier\,\orcidlink{0000-0001-6691-3695}}
\email{leonardo.chataignier@ehu.eus}
\affiliation{Department of Physics and EHU Quantum Center, University of the Basque Country UPV/EHU, Barrio Sarriena s/n, 48940 Leioa, Spain}

\begin{abstract}
We revisit the concept of phase time, which has been previously proposed as a solution to the problem of time in quantum gravity. Concretely, we show how the geometry of configuration space together with the phase of the wave function of the universe can, under certain conditions, lead to the definition of a positive-definite squared norm from which probabilities can be defined. The norm is conserved under time evolution, and we obtain a version of Ehrenfest's theorem that is analogous to the one in ordinary quantum mechanics. We address how the present approach avoids some difficulties that were encountered in previous attempts at defining an Ehrenfest theorem relative to phase time in the context of relativistic quantum mechanics and canonical quantum gravity, and how it is connected to a notion of gauge fixing the inner product. We conclude with a brief outlook.
\end{abstract}

\maketitle

\section{Introduction}
Greensite and Padmanabhan independently suggested that the phase of the wave function of the universe should be used to define a notion of time in canonical quantum gravity \cite{Greensite,Padmanabhan,Greensite1,Greensite2}, thereby presenting a possible solution to the problem of time. This problem concerns the description of a diffeomorphism-invariant quantum dynamics, and it inevitably follows from the Hamiltonian constraint(s), also known as the Wheeler--DeWitt equation(s) \cite{DeWitt,Kiefer:book}, in canonical quantum gravity. Their work followed previous developments in the so-called ``semiclassical'' or Born--Oppenheimer approach to quantum gravity, in which time is defined from a part of the phase of the wave function that is computed from a Wentzel--Kramers--Brillouin (WKB) expansion for the gravitational field. This motivated the term ``WKB time'' for this construction \cite{Zeh}. More generally, the WKB expansion is performed for ``heavy'' fields associated with a large mass scale, such as the Planck mass (see \cite{LapRuba,Banks,Brout1,Zeh,Venturi,KieferSingh,KieferRev} for references that established this Born--Oppenheimer approach prior to and around the time of the work of Greensite and Padmanabhan). It was then natural to extend this definition of time, as Padmanabhan and Greensite did, to the total phase of the wave function of the universe without the need to perform a WKB expansion. Following \cite{Brotz}, we refer to this type of definition as ``phase time.''

A central feature of the Greensite--Padmanabhan approach was the requirement that a so-called ``Ehrenfest principle'' be satisfied. This meant that a version of Ehrenfest's theorem should be valid, despite the fact that physical states must be annihilated by the Hamiltonian operator (making them stationary states). This was to be achieved by expressing the evolution relative to phase time. Their construction, however, suffered from some limitations discussed by Greensite \cite{Greensite2} and by Brotz and Kiefer \cite{Brotz}: most notably, one needed extra assumptions in order to obtain the Ehrenfest-theorem dynamics of the canonical momenta, which were not readily obtained. There have also been other discussions of Ehrenfest's theorem or, more generally, of the dynamics of expectation values in the context of toy models of quantum gravity that did not make explicit use of the phase-time concept (see, for example, \cite{Misner57,Squires}).

Here, we revisit the idea of phase time in general, and we show how one can obtain a version of Ehrenfest's theorem that avoids the difficulties of the previous construction, provided some simple assumptions are made about the wave function and the geometry of configuration space. We focus on minisuperspace (i.e., mechanical) models in order to keep the discussion focused on the problem of time and its phase-time solution, and field-theoretic regularization issues are not discussed, even though we offer some remarks on the full theory. The approach described here is also useful to general Born--Oppenheimer and WKB-time systems, but the details of this application will be discussed elsewhere, in order to keep the current discussion focused on the essential issue of Ehrenfest's theorem and the comparison with the previous works by Greensite and Padmanabhan.

The article is divided into six sections. In Sec.~\ref{sec:nonrel}, we cast the Ehrenfest theorem of ordinary (nonrelativistic) quantum mechanics into a ``polar form,'' which will be useful for comparison with the phase-time Ehrenfest theorem to be established in Sec.~\ref{sec:rel}. There, we examine (``first-quantized'') relativistic quantum theory, and we show how the phase of the wave function can be related to a foliation of configuration space, with which a positive-definite norm can be defined. We emphasize how this norm is connected to a version of Faddeev--Popov gauge fixing. The ensuing probability distribution is conserved in time, and it can be thought of as referring to gauge-independent quantities. From this, a version of Ehrenfest's theorem follows in polar form. In Sec.~\ref{sec:qg}, we discuss how this formalism could be translated to a canonical quantum field theory of gravity. Finally, in Sec.~\ref{sec:eg}, we illustrate the approach presented here with a simple cosmological application, and we conclude with an outlook to further research in Sec.~\ref{sec:conclusions}.

\section{\label{sec:nonrel}Nonrelativistic quantum mechanics}
The central equation of ordinary nonrelativistic\footnote{Here, we use a somewhat generalized terminology and refer to theories as ``nonrelativistic'' when they single out a preferred class of time variables, so as not to display a nontrivial form of local time-reparametrization invariance (see comments in Sec. \ref{sec:rel}). Such ``nonrelativistic'' theories may of course be relativistic in the Galilean sense, for example.} quantum mechanics is the time-dependent Schr\"odinger equation:
\eq{\label{eq:schro}
\I\hbar\del{}{\tau}\Psi = \hat{\mathcal{H}}\Psi \ .
}
For concreteness, we assume that the configuration-space representation of $\hat{\mathcal{H}}$ has the form:
\eq{\label{eq:nonrel-H}
\hat{\mathcal{H}}&:= \hat{H}-\left(\I\hbar\del{}{\tau}\log|g|^{\frac14}\right) \ , \\
\hat{H}&:= -\frac{\hbar^2}{2}\nabla^2+V(q) \ ,
}
where $V(q)$ is a potential that depends on the configuration point $q$, and $\nabla^2$ is the Laplace--Beltrami operator,
\eq{\label{eq:nabla}
\nabla^2:=\frac{1}{\sqrt{|g|}}\del{}{q^{\mu}}\left(\sqrt{|g|}g^{\mu\nu}\del{}{q^{\nu}}\right) \ ,
}
on the $(d+1)$-dimensional configuration-space manifold with metric coefficients $g_{\mu\nu}$, inverse metric coefficients $g^{\mu\nu}$, and metric determinant $g$.\footnote{The signature of the metric may be pseudo-Riemannian (Lorentzian).} Notice that a summation over repeated indices is tacitly performed in Eq. \eref{nabla} and throughout the article. As we allow the configuration-space metric to have an explicit time dependence, the last term in Eq. \eref{nonrel-H} is necessary to guarantee a unitary dynamics (see, e.g., \cite{DeWitt57}) with respect to the inner product $\braket{\cdot|\cdot}$, relative to which the operator $\hat{H} = -\hbar^2\nabla^2/2+V$ is self-adjoint. Indeed, the inner product $\braket{\cdot|\cdot}$ in this case is
\eq{\label{eq:nonrel-IP}
\hspace{-0.2cm}\braket{\Psi_2(\tau)|\Psi_1(\tau)}_\tau \!:=\!\! \int\!\D^{d+1} q \sqrt{|g(\tau;q)|}\Psi_2^*(\tau;q)\Psi_1(\tau;q) ,
}
and $\braket{\Psi_2(\tau)|\Psi_1(\tau)}_{\tau}$ is conserved with respect to time if $\Psi_{1,2}(\tau;q)$ are two solutions to Eq. \eref{schro}.

Assuming that $\sqrt{|g|}\Psi_2^*\Psi_1$ vanishes at the boundaries of configuration space, the momentum operator that is self-adjoint relative to the inner product in Eq. \eref{nonrel-IP} is \cite{DeWitt57}
\eq{\label{eq:momentum}
\hat{p}_{\mu}:= \frac{\hbar}{\I|g|^{\frac14}}\del{}{q^{\mu}}|g|^{\frac14} \ ,
}
and it obeys the commutation relation $[q^{\mu},p_{\nu}] = \I\hbar\delta^{\mu}_{\nu}$. A general complex solution, with polar decomposition
\eq{\label{eq:polar}
\Psi(\tau;q) = R(\tau;q)\exp\left[\frac\I\hbar S(\tau;q)\right] \ ,
}
has the squared norm
\eq{
\braket{\Psi(\tau)|\Psi(\tau)}_{\tau} = \int\D^{d+1} q\, \sqrt{|g(\tau;q)|}\, R^2(\tau;q) \ ,
}
and it leads to the conserved probability distribution
\eq{\label{eq:Born}
p_{\tau;\Psi}(q) := \frac{\sqrt{|g(\tau;q)|}\, R^2(\tau;q)}{\int\D^{d+1} q'\, \sqrt{|g(\tau;q')|}\, R^2(\tau;q')} \ .
}
The expectation value of the configuration point is
\eq{\label{eq:av-q}
\!\!\braket{q^{\mu}(\tau)} := \int\D^{d+1}q\, q^{\mu}\,p_{\tau;\Psi}(q)= \frac{\braket{\Psi(\tau)|q^{\mu}\Psi(\tau)}_\tau}{\braket{\Psi(\tau)|\Psi(\tau)}_\tau} \,,
}
whereas the expectation value of momenta can be defined as
\eq{\label{eq:av-p}
\braket{p_{\mu}(\tau)} &:= \int\D^{d+1}q\, \del{S}{q^{\mu}}\, p_{\tau;\Psi}\\
&= \frac{\braket{\Psi(\tau)|\hat{p}_{\mu}\Psi(\tau)}_\tau}{\braket{\Psi(\tau)|\Psi(\tau)}_\tau}\\
& \ \ \ +\frac{\I\hbar\int\D^{d+1}q\, \del{}{q^{\mu}}\left(\sqrt{|g|}R^2\right)}{2\braket{\Psi(\tau)|\Psi(\tau)}_\tau}
}
because of
\eq{\label{eq:momentum-polar}
\frac{\hat{p}_{\mu}\Psi}{\Psi} = \del{S}{q^{\mu}}+\frac{\hbar}{2\I}\del{}{q^{\mu}}\log\left(\sqrt{|g|}R^2\right) \ .
}
The last term in Eq. \eref{av-p} vanishes if the solution is normalizable with $\sqrt{|g|}R^2$ vanishing at the boundaries of configuration space. This then yields the usual expression.

The dynamics of these expectation values can be computed by using Eq. \eref{schro} or its polar decomposition, which leads to the equations:
\eq{\label{eq:polar-schro1}
-\del{S}{\tau} =\frac12g^{\mu\nu}\del{S}{q^{\mu}}\del{S}{q^{\nu}}+\V(q)
}
where
\eq{\label{eq:q-potential}
\V(q) := V(q)-\frac{\hbar^2}{2}\frac{\nabla^2R}{R} \ ,
}
and
\eq{\label{eq:polar-schro2}
\del{}{\tau}\left(\sqrt{|g|}R^2\right) &=-\del{}{q^{\mu}}\left(\sqrt{|g|}R^2g^{\mu\nu}\del{S}{q^{\nu}}\right) \ .
}
As is well known, whereas Eq. \eref{polar-schro1} is a quantum version of the Hamilton--Jacobi equation, the equation \eref{polar-schro2} corresponds to a continuity equation for the probability distribution $\sqrt{|g|}R^2$ [cf. Eq. \eref{Born}] if the solution is normalized, i.e., $\braket{\Psi(\tau)|\Psi(\tau)}_\tau = 1$, which we assume from now on for simplicity.

Using Eq. \eref{schro} together with Eqs. \eref{av-q} and \eref{av-p}, we obtain
\eq{\label{eq:Ehrencom1}
\del{}{\tau}\braket{q^{\mu}(\tau)} = \left<\!\Psi(\tau)\left|\frac\I\hbar\left[\hat{H},q^{\mu}\right]\right.\!\!\Psi(\tau)\!\right>_{\tau} ,
}
and
\eq{\label{eq:Ehrencom2}
\del{}{\tau}\braket{p_{\mu}(\tau)} = \left<\!\Psi(\tau)\left|\frac\I\hbar\left[\hat{H},\hat{p}_{\mu}\right]\right.\!\!\Psi(\tau)\!\right>_{\tau} .
}
These equations are the results of Ehrenfest's theorem for the particular case of the configuration point and momenta operators. Now notice that, for a real configuration-space function $f\equiv f(q)$, the commutator
\eq{\label{eq:com}
[f,\hat{H}] = \hbar^2g^{\mu\nu}\del{f}{q^{\mu}}\del{}{q^{\nu}}+\frac{\hbar^2}{2}\nabla^2f
}
leads to the identity
\eq{\label{eq:identity-com}
\sqrt{|g|}\Psi^*\frac{\I}{\hbar}[\hat{H},f]\Psi &= \sqrt{|g|}R^2\,g^{\mu\nu}\del{f}{q^{\mu}}\del{S}{q^{\nu}}\\
&\ \ \ -\frac{\I\hbar}{2}\del{}{q^{\mu}}\left(\sqrt{|g|}R^2\,g^{\mu\nu}\del{f}{q^{\nu}}\right)\,,
}
which implies
\eq{\label{eq:Ehren1}
\del{}{\tau}\braket{q^{\mu}(\tau)} = \int\D^{d+1}q\,\sqrt{|g|}R^2\, g^{\mu\nu}\del{S}{q^{\nu}} \ ,
}
because the last term in Eq. \eref{identity-com}, upon integration, yields a boundary term that vanishes if $\sqrt{|g|}R^2$ is zero at the boundaries for the normalizable state $\Psi$.

Similarly, using Eqs. \eref{schro}, \eref{nonrel-H}, \eref{momentum}, \eref{polar},  and \eref{polar-schro2}, as well as the self-adjointness of $\hat{H}$, we can write
\eq{\label{eq:Ehren2-pre}
&\del{}{\tau}\braket{p_{\mu}(\tau)}\\
&= -\frac{2}{\hbar}\mathrm{Im}\braket{\hat{H}\Psi(\tau)|\hat{p}_{\mu}\Psi(\tau)}_{\tau}\\
&= 2\hbar\mathrm{Im}\int\D^{d+1}q\!\left(\del{}{\tau}|g|^{\frac14}\Psi\right)^{\!*}\!\!\left(\del{}{q^{\mu}}|g|^{\frac14}\Psi\right)\\
&= -\int\D^{d+1}q\, \del{S}{\tau}\del{}{q^{\mu}}\sqrt{|g|}R^2\\
&\ \ \ +\int\D^{d+1}q\,\del{S}{q^{\mu}}\del{}{\tau}\sqrt{|g|}R^2\\
&= \int\D^{d+1}q\, \sqrt{|g|}R^2\, \del{^2S}{q^{\mu}\partial\tau}\\
&\ \ \ +\int\D^{d+1}q\,\sqrt{|g|}R^2\,g^{\rho\sigma}\del{S}{q^{\rho}}\del{^2S}{q^{\sigma}\partial q^{\mu}}\,,
}
where we used integration by parts and the assumption that boundary terms vanish as $\sqrt{|g|}R^2$ is zero at the boundaries. By using Eq. \eref{polar-schro1} in Eq. \eref{Ehren2-pre}, we finally obtain
\eq{\label{eq:Ehren2}
&\del{}{\tau}\braket{p_{\mu}(\tau)}\\
&= -\int\D^{d+1}q\,\sqrt{|g|}R^2\, \left(\frac12\del{g^{\rho\sigma}}{q^{\mu}}\del{S}{q^{\rho}}\del{S}{q^{\sigma}}+\del{\V}{q^{\mu}}\right) .
}
The equations \eref{Ehren1} and \eref{Ehren2} are thus equivalent to Eqs. \eref{Ehrencom1} and \eref{Ehrencom2} but they make use of the polar decomposition given in Eq. \eref{polar}.\footnote{In fact, we can obtain Eqs. \eref{Ehren1} and \eref{Ehren2} directly from Eq. \eref{av-q} and the first line of Eq. \eref{av-p} by using Eqs. \eref{polar-schro1} and \eref{polar-schro2}.} This ``polar form'' of the Ehrenfest equations will be of central interest in the relativistic case to be discussed next.

If $\Psi(\tau;q)$ is a wave packet that is very narrowly peaked during some time interval $\Delta\tau$, it is straightforward to see that the classical dynamics is recovered from Eqs. \eref{Ehren1} and \eref{Ehren2} during $\Delta\tau$. More precisely, if $\sqrt{|g|}R^2$ approximates a Dirac delta distribution centered at $q = q_0$ at an instant $\tau$, then Eqs. \eref{Ehren1} and \eref{Ehren2} lead to
\eq{\label{eq:Ehren1c}
\del{}{\tau}\braket{q^{\mu}(\tau)} &\simeq \left(g^{\mu\nu}\del{S}{q^{\nu}}\right)_{q=q_0}\\
&=\{q^{\mu},H\}_{q=q_0,p=\partial S/\partial q|_{q = q_0}}
}
and
\eq{\label{eq:Ehren2c}
\del{}{\tau}\braket{p_{\mu}(\tau)}\!&\simeq\! -\left(\frac12\del{g^{\rho\sigma}}{q^{\mu}}\del{S}{q^{\rho}}\del{S}{q^{\sigma}}+\del{\V}{q^{\mu}}\right)_{q=q_0}\\
&= \{p_{\mu},H\}_{q=q_0,p=\partial S/\partial q|_{q = q_0}}
}
at the instant $\tau$, with $H = g^{\mu\nu}p_{\mu}p_{\nu}/2+\V$ being the Hamiltonian phase-space function, and $\{\cdot,\cdot\}$, the Poisson bracket. In these equations, the momenta are given by the gradient of $S$.

Depending on the form of the probability distribution, the mean values $q_0$ may depend on $\hbar$. Likewise, depending on the phase, the momenta $p = \partial S/\partial q \equiv \nabla S$ may also have an $\hbar$ dependence. If both $R$ and $S$ admit an expansion in a series of powers of $\hbar$, with lowest order $\hbar^0$, then the Eqs.~\eref{Ehren1c} and~\eref{Ehren2c} reduce to classical equations that are independent of $\hbar$ at the lowest order. Indeed, up to the lowest order in $\hbar$, the phase factor $S$ is a solution to the Hamilton--Jacobi equation, whereas $\V\simeq V|_{\hbar = 0}$ [cf. Eqs. \eref{polar-schro1} and \eref{q-potential}], and the mean values calculated from $R$ would be independent of $\hbar$ at the lowest order. With this, the expectation values $\braket{q^{\mu}(\tau)}$ and $\braket{p_{\mu}(\tau)}$ approximate a classical solution, as long as the corresponding probability distribution remains narrowly peaked.

We now turn to the relativistic case in order to discuss how a similar result can be established.

\section{\label{sec:rel}Relativistic quantum mechanics}
A central feature of (special) relativistic mechanics and gravitation alike is the absence of an absolute, external time parameter $\tau$ that governs the dynamics. Rather, the evolution is to be described relative to time variables defined from the physical degrees of freedom themselves and that keep track of, for example, an observer's proper time or an affine parameter along a particle's worldline. All the different choices of such ``internal'' time variables are on equal footing, which corresponds to the statement that the theory is time-reparametrization invariant.\footnote{For (special) relativistic particles, this is equivalent to observing that the theory is symmetric under worldline diffeomorphisms (that are possibly phase-space dependent), whereas it is symmetric under spacetime diffeomorphisms (that are possibly field dependent) in general relativity.}

In this way, the corresponding quantum theory must also be independent of an external time $\tau$ if it is to maintain the same local reparametrization symmetry. We can construct such a theory from the nonrelativistic formulae in Sec. \ref{sec:nonrel} by demanding that all quantities, including the configuration-space metric and the wave function, be $\tau$ independent. Thus, instead of Eqs. \eref{schro} and \eref{nonrel-H}, we obtain the time-independent Schr\"odinger equation\footnote{This defines the relativistic quantum mechanics of a spin 0 field. As it is mechanical, it is a ``first-quantized'' theory, as opposed to a quantum field theory (sometimes referred to as ``second quantized'').}
\eq{\label{eq:KG}
0 = \hat{H}\Psi = -\frac{\hbar^2}{2}\nabla^2\Psi+V(q)\Psi \ ,
}
which corresponds to the Klein--Gordon equation on a (possibly curved) background if the configuration space is a pseudo-Riemannian manifold.\footnote{Such Klein--Gordon equations are often used in toy models of quantum cosmology \cite{Kiefer:book}, in which case $\Psi$ can be seen as the ``wave function of the universe.'' See the example in Sec.~\ref{sec:eg}.} We will assume that this is the case, as the configuration spaces of both relativistic particles (which travel through spacetime) and of general relativity have Lorentzian signatures $(-,+,\ldots,+)$.\footnote{The Lorentzian signature of the metric of the configuration space of general relativity is \emph{a priori} unrelated to causality in spacetime, which is rather associated with the Lorentzian signature of the metric of spacetime (not configuration space). The spacetime metric (or more precisely, some of its components) is used to define points in the configuration space of general relativity \cite{DeWitt,Kiefer:book}.} As a result, the squared norm of configuration-space vectors may be positive, zero, or negative. To avoid confusion with the analogous spacetime concepts, we refer to such vectors as positivelike, null, or negativelike, respectively, instead of spacelike, null or lightlike, and timelike.

By using the polar decomposition given in Eq. \eref{polar}, we can decompose Eq. \eref{KG} into the equations:
\eq{\label{eq:polar-KG1}
\frac12g^{\mu\nu}\del{S}{q^{\mu}}\del{S}{q^{\nu}}+\V(q) = 0
}
and
\eq{\label{eq:polar-KG2}
\del{}{q^{\mu}}\left(\sqrt{|g|}R^2g^{\mu\nu}\del{S}{q^{\nu}}\right) = 0 \ ,
}
which are simply Eqs. \eref{polar-schro1} and \eref{polar-schro2} restricted to the case in which all quantities are independent of $\tau$. In particular, $\V(q)$ is still given by Eq. \eref{q-potential}.

\subsection{\label{sec:IP}Inner product}
While $\hat{H}$ is still self-adjoint with respect to the inner product given in Eq. \eref{nonrel-IP} (but now with all quantities being independent of $\tau$), there is no notion of unitary evolution with respect to the external time $\tau$ because this parameter is no longer present.\footnote{Moreover, if zero is in the continuous spectrum of the operator $\hat{H}$, then the solutions to Eq. \eref{KG} will not be normalizable with respect to the inner product in Eq. \eref{nonrel-IP}, and they will not define probability distributions as in Eq. \eref{Born}. Although the diverging norms of solutions can be regularized via the Rieffel induction or refined algebraic quantization and group averaging procedures (see, e.g., \cite{Hartle}), we do not consider these formalisms here, as they often rely on the spectral analysis of $\hat{H}$ that we will not discuss. Rather, we focus solely on properties of the configuration-space manifold and the wave function $\Psi(q)$, and we discuss under what conditions they may lead to the Ehrenfest dynamics as in the nonrelativistic case. See also the discussion in Sec.~\ref{sec:conclusions}.} However, the form of the Hamiltonian guarantees that a conserved current exists:
\eq{\label{eq:KG-cont}
&\del{}{q^{\mu}}\left(\sqrt{|g|}j^{\mu}[\Psi_2,\Psi_1]\right) =0 \ , \\
&j^{\mu}[\Psi_2,\Psi_1] := -\frac{\I\hbar}{2} g^{\mu\nu}\left(\Psi_2^*\del{\Psi_1}{q^{\nu}}-\Psi_1\del{\Psi_2^*}{q^{\nu}}\right) \ .
}
This is the well-known Klein--Gordon current. Notice that it obeys the continuity equation \eref{KG-cont} in configuration space and without an external $\tau$, and thus its conservation can be interpreted with respect to an internal time (if one can be defined). This current is not often considered because it leads to the Klein--Gordon inner product
\eq{\label{eq:KGprod}
(\Psi_2,\Psi_1):=\int_{\Sigma}\D^{d}\Sigma_{\mu}\,j^{\mu}[\Psi_2,\Psi_1] \ .
}
Although this inner product is conserved in the sense of being independent of the choice of positivelike hypersurface $\Sigma$ (with volume element $\D^{d}\Sigma_{\mu}$), it is generally regarded as inadequate because it is indefinite; i.e., it allows states with positive, zero, and negative squared norm. This indefiniteness would cause difficulties with the definition of non-negative probability distributions as in Eq. \eref{Born}. Clearly, only states with nonzero norm can, up to a state-dependent sign, be used to define probability distributions.\footnote{This is the case unless one is prepared to depart, to a greater or lesser extent, from the established formalism of quantum mechanics or probability theory.} We will see that, under certain conditions, such states do indeed lead to well-defined probabilities with an associated Ehrenfest dynamics.

Notice that the Klein--Gordon current $j^{\mu}[\Psi]:=j^{\mu}[\Psi_2 = \Psi, \Psi_1 = \Psi] = R^2g^{\mu\nu}\partial{S}/\partial{q^{\nu}}$, for which Eq. \eref{polar-KG2} coincides with the first of Eqs. \eref{KG-cont}, implies that states for which $S$ is constant across configuration space (and, in particular, real states, for which $S$ is identically zero) have zero norm. Therefore, we exclude the case in which the phase factor $S$ is constant.

\subsection{\label{sec:fol}Configuration-space foliation}
Let us then consider the case in which $S$ is a nontrivial, nonconstant solution to Eqs. \eref{polar-KG1} and \eref{polar-KG2}. Its gradient can be used to define the configuration-space vector field \cite{Kuchar,Isham,Chataignier:2019,Chataignier:2022,Chataignier:2023}
\eq{\label{eq:phase-grad}
X = \E(q) g^{\mu\nu}(q)\del{S}{q^{\mu}}\del{}{q^{\nu}} \ ,
}
where $\E(q)$ is a nonvanishing configuration-space scalar that can be arbitrarily chosen. For convenience, we choose it to be always positive.\footnote{We will later see what the dynamical significance of $\E(q)$ is. In Sec.~\ref{sec:gf}, we comment on the fact that the choice of $\E(q)$ is subject to admissibility conditions rather than being completely arbitrary. See also the appearance of $\E(q)$ in the equations of motion in the classical limit in Sec.~\ref{sec:ehrenfest}.} The solutions $\varphi(q)$ to the differential equation $X\varphi = 1$ can then induce a foliation of configuration space.

More precisely, besides being a pseudo-Riemannian manifold, let us make the further assumption that the $(d+1)$-dimensional configuration space can be foliated into a one-parameter family of positivelike leaves (i.e., hypersurfaces with negativelike normal vectors), which are equal to the level sets of a smooth scalar $\varphi(q)$. We choose this scalar to satisfy $X\varphi = 1$ with the additional assumption that the gradient of $\varphi$ is not equal to zero anywhere. In general, $\varphi(q)$ will only be defined in terms of the configuration-point coordinates $q^{\mu}$ in local regions of configuration space (i.e., it may not be possible to define it globally in terms of only one set of coordinates), so that it is preferable to take $X$ as the fundamental quantity that defines the foliation.

The unit normal vector to the leaf defined by $\varphi(q)=s$ has the components $n^{\mu}=-Ng^{\mu\nu}\partial_{\nu}\varphi$, where $N$ is a normalization function [$N=\pm(-g^{\mu\nu}\partial_{\mu}\varphi\partial_{\nu}\varphi)^{-\frac12}$; it is sufficient to choose the positive sign so that $N>0$],\footnote{In the usual Arnowitt--Deser--Misner (ADM) decomposition of general relativity, $N$ plays the role of the lapse function (see, e.g., \cite{Gourgoulhon}). Here, however, since the foliation concerns the configuration space, $N$ is \textit{a priori} unrelated to the spacetime lapse function.} and the vector field $X$ generates translations in the leaf parameter $s$ because $\varphi(q+\epsilon X) = \varphi(q)+\epsilon X\varphi+\Ob(\epsilon^2) = s + \epsilon+\Ob(\epsilon^2)$. In this way, due to $X\varphi = 1$, the components of $X$ can be written as $X^{\mu} = Nn^{\mu}+\tilde{\beta}^{\mu}$, where $\tilde{\beta}^{\mu}\partial_{\mu}\varphi=0$.\footnote{In the ADM decomposition, $\tilde{\beta}^{\mu}$ corresponds to the components of the spacetime shift vector \cite{Gourgoulhon}, whereas here it is the analogous quantity in configuration space.}

The foliation given by $\varphi(q)=s$ for all possible values of $s$ can be used to define new configuration-space coordinates $q^{\mu}\mapsto x^{\nu} = (s,x^i)$, where $x^i$ are coordinates on the leaves, which lead to the orthogonality relation $n_{\mu}\partial_i q^{\mu}=0$. As the vector with components $\tilde{\beta}^{\mu}$ is tangent to the leaves, we can write $\tilde{\beta}^{\mu}\mapsto\beta^{\nu}=\delta^{\nu}_i\beta^i$ in the new coordinates (i.e., the component along $\partial/\partial x^i$ is $\beta^i$, whereas the component along $\partial/\partial s$ is zero). The new metric components read
\eq{\label{eq:decomp-metric}
\tilde{g}_{00} &= g_{\mu\nu}\del{q^{\mu}}{s}\del{q^{\nu}}{s} = X^{\mu}X_{\mu} =  \beta^i\beta_i-N^2\ ,\\
\tilde{g}_{0i} &= g_{\mu\nu}\del{q^{\mu}}{s}\del{q^{\nu}}{x^i} = X_{\mu}\del{q^{\mu}}{x^i} = \tilde{\beta}_{\mu}\del{q^{\mu}}{x^i} = \beta_i \ , \\
\tilde{g}_{ij} &=g_{\mu\nu}\del{q^{\mu}}{x^i}\del{q^{\nu}}{x^j} =: h_{ij} \ ,
}
where $\beta_i = h_{ij}\beta^j$.

The equations \eref{decomp-metric} present the well-known decomposition of a metric relative to coordinates adapted to the (positivelike) foliation \cite{Gourgoulhon,Kiefer:book}. Likewise, the inverse metric has components
\eq{\label{eq:decomp-inv-metric}
\tilde{g}^{00} &= -\frac{1}{N^2}\ ,\ 
\tilde{g}^{0i} = \frac{\beta^i}{N^2} \ , \ 
\tilde{g}^{ij} = h^{ij}-\frac{\beta^i\beta^j}{N^2}\ ,
}
with $h^{ij}$ defined by $h^{ij}h_{jk} = \delta^i_k$, and the metric determinant satisfies $\tilde{g} := \det \tilde{g}_{\mu\nu} = -N^2 h$, with $h:=\det h_{ij}$.

Notice that, because of Eq. \eref{polar-KG1}, the norm of $X$ satisfies
\eq{\label{eq:norm-X}
X^{\mu}X_{\mu} = \E^2g^{\mu\nu}\del{S}{q^{\mu}}\del{S}{q^{\nu}} = -2\E^2\V \ .
}
Thus, $X$ need not be negativelike, so that $\tilde{g}_{00}$ could be positive, zero, or negative depending on the sign of $\V$. In the particular case in which $\V$ is positive everywhere, $\tilde{g}_{00}$ is everywhere negative and $X$ is everywhere negativelike. 

We can establish some further useful identities for this configuration-space foliation. First, by comparing Eqs. \eref{phase-grad} and \eref{decomp-metric}, we find
\eq{\label{eq:S-N-beta}
\beta_i &= \E \del{S}{x^i} \ , \\
N^2 &=\beta^i\beta_i+ 2\E^2\V \ ,
}
and
\eq{\label{eq:momentum-time}
\del{S}{s} &= XS =\E g^{\mu\nu}\del{S}{q^{\mu}}\del{S}{q^{\nu}} = -2\E\V = \frac{\tilde{g}_{00}}{\E}\ .
}
Second, we can use the coordinates adapted to the foliation to compute the Laplace--Beltrami operator acting on the phase factor to find \cite{Chataignier:2019,Chataignier:2022,Chataignier:2023}:
\eq{\label{eq:nabla2S}
\nabla^2S = \frac{1}{N\sqrt{h}}\del{}{s}\left(\frac{N\sqrt{h}}{\E}\right)\ ,
}
where we also used Eqs. \eref{S-N-beta}.

Finally, due to the similarity of Eq.~\eref{polar-KG1} with the classical Hamilton--Jacobi equation, the action of $X$ on the configuration point $q^{\mu}$ and on the gradient components $\partial S/\partial q^{\mu}$ can also be written in terms of Poisson brackets $\{\cdot,\cdot\}$. Indeed, we find
\eq{\label{eq:X-flow-q}
Xq^{\mu} = \E g^{\mu\nu}\del{S}{q^{\nu}} = \E\{q^{\mu}, H\}_{p = \partial S/\partial q} \ ,
}
with $H = g^{\mu\nu}p_{\mu}p_{\nu}/2+\V$, which vanishes when $p = \partial S/\partial q\equiv\nabla S$.\footnote{Notice that the Poisson bracket is computed before $H = 0$ is enforced. This is the standard procedure for constrained Hamiltonian systems, as was emphasized by Dirac \cite{Dirac1,Dirac2,Dirac3} for example.} Similarly, due to Eq. \eref{polar-KG1}, we can use the identity
\eq{
g^{\rho\sigma}\del{S}{q^{\rho}}\del{^2S}{q^{\sigma}\partial q^{\mu}}\! &= \frac12\del{}{q^{\mu}}\left(g^{\rho\sigma}\del{S}{q^{\rho}}\del{S}{q^{\sigma}}\right)\\
&\ \ \ -\frac12\del{g^{\rho\sigma}}{q^{\mu}}\del{S}{q^{\rho}}\del{S}{q^{\sigma}}\\
& = -\del{\V}{q^{\mu}}-\frac12\del{g^{\rho\sigma}}{q^{\mu}}\del{S}{q^{\rho}}\del{S}{q^{\sigma}} \,,\label{eq:p-X-id}
}
to find
\eq{\label{eq:X-flow-p}
X\del{S}{q^{\mu}} = \E\{p_{\mu},H\}_{p = \partial S/\partial q} \ .
}
The connection of Eqs.~\eref{X-flow-q} and \eref{X-flow-p} with the classical trajectories will be discussed in Sec.~\ref{sec:ehrenfest} in analogy to the nonrelativistic Eqs.~\eref{Ehren1c} and \eref{Ehren2c}.\footnote{Notice that Eqs.~\eref{X-flow-q} and \eref{X-flow-p} can be used to define trajectories beyond the classical limit, which follow from the flow of the configuration-space vector field $X$. As this vector field is defined from the gradient of the phase factor $S$, these trajectories coincide with the ones considered in the pilot-wave or de Broglie--Bohm approach to quantum theory (see, for instance, \cite{PW1,PW2}). In this approach, one assumes that such trajectories exist in the sense that the quantum system actually follows one of them. In the present article, however, it is irrelevant whether such trajectories are followed or not. The fact of the matter is that, given a solution to Eq.~\eref{KG} (with a nontrivial phase factor $S$), $X$ and its flow can be defined as mathematical constructs in configuration space. We will see in Sec.~\ref{sec:ehrenfest} how these constructs are then related to the classical trajectories via the Ehrenfest theorem.}

\subsection{\label{sec:probs-gen}Probabilities}
With the state-dependent configuration-space foliation just defined, the conservation of $j^{\mu}[\Psi]$ leads to\footnote{A particular case of Eq.~\eref{conserve-integrand} was also obtained by Brotz and Kiefer \cite{Brotz} using specific assumptions that will be discussed in Sec.~\ref{sec:GP}.} [cf. Eqs. \eref{polar-KG2} and \eref{nabla2S}]
\eq{\label{eq:conserve-integrand}
0 = \frac{1}{\E}\del{R^2}{s}+R^2\nabla^2S = \frac{1}{N\sqrt{h}}\del{}{s}\left(\frac{N\sqrt{h} R^2}{\E}\right) \ ,
}
and the Klein--Gordon inner product yields the squared norm
\eq{\label{eq:fol-norm}
(\Psi,\Psi) = \int_{\Sigma}\D^{d}x\ \frac{N\sqrt{h} R^2}{\E} \ ,
}
where $\Sigma$ is one of the leaves, and we used Eqs. \eref{decomp-inv-metric} and \eref{S-N-beta}. Because of Eq. \eref{conserve-integrand}, we see that the squared norm is independent of $s$, $\partial(\Psi,\Psi)/\partial s = 0$, and so it is conserved if $s$ is taken as an internal time parameter. In this case, the $x$ coordinates can be seen as the independent variables that determine the possible initial values of the configuration point, $q^{\mu}(s = s_0,x)$.

From now on, we assume that the state $\Psi$ has been suitably normalized so that $(\Psi,\Psi) = 1$ for all values of $s$. Since $N>0$ and $\E > 0$, the squared norm given in Eq. \eref{fol-norm} is positive definite for a suitable square-integrable amplitude $R(q(s,x))$, and thus it can be used to define an $s$-independent probability density:
\eq{\label{eq:probs}
p_{\Psi}(x) := \frac{N\sqrt{h} R^2}{\E} \ .
}
The assumptions that led to this result were
\begin{itemize}
    \item The wave function is a nontrivial solution to Eq. \eref{KG}, and its phase has a nontrivial gradient that allows for a nonzero Klein--Gordon squared norm.
    \item It is possible to foliate the pseudo-Riemannian configuration space into a one-parameter family of positivelike leaves defined from the level sets of $\varphi$, with the choices $N>0, \E>0$. These choices imply that the Klein--Gordon squared norm of the state is, in fact, positive definite.  The interpretation of $\E$ will be discussed below.\footnote{Furthermore, in the example discussed in Sec.~\ref{sec:eg}, we will see how the restriction $\E>0$ reduces to the the restriction to positive frequencies for a free, massless relativistic particle in Minkowski spacetime.}
\end{itemize}
It is worthwhile to emphasize that these assumptions may of course not be satisfied in certain situations. For instance, depending on the precise form of $\hat{H}$ and the potential $V(q)$ in Eq.~\eref{KG}, it may be difficult to find a nontrivial solution with a nontrivial phase gradient in a straightforward manner. Moreover, it may not be possible to globally foliate the configuration space (e.g., if it is not a globally hyperbolic manifold). There may also be no choice of $\E$ in Eq.~\eref{phase-grad} that leads to a globally well-defined flow for the vector field $X$, and, as we have already mentioned, it may not be possible to define $\varphi(q)$ globally. These limitations indicate that a notion of phase time may be inexorably an approximate one, and that it should be subsumed by a more fundamental formalism, such as, perhaps, the quantization of the Becchi--Rouet--Stora--Tyutin (BRST) cohomology (see \cite{HT:book} and the comments in Sec.~\ref{sec:conclusions}). Since our interest in this article is in showing how the Ehrenfest theorem can be obtained in the context of the phase-time construction, and how it avoids the difficulties encountered in the previous approach due to Greensite and Padmanabhan, we assume in what follows that there are no obstructions to defining the probability density as in Eq.~\eref{probs}.

\subsection{\label{sec:phase-times}Phase times}
As the internal time $s$ is defined from the level sets of the scalar $\varphi$, which is a solution to $X\varphi = 1$, and $X$ is proportional to the gradient of the phase factor $S$, we can refer to $s$ as the phase time. Notice, however, that the phase time can, in fact, coincide with different time variables depending on the arbitrary choice of $\E$. Indeed, using Eq. \eref{phase-grad}, we can rewrite the equation $X\varphi = 1$ as
\eq{\label{eq:E-phase-grad}
\frac{1}{\E} = g^{\mu\nu}\del{S}{q^{\mu}}\del{\varphi}{q^{\nu}} \ .
}
This form of the equation makes it clear that one can first choose the state [viz. $S$, a solution to Eq. \eref{polar-KG1}] together with $\E$ and solve for $\varphi$, or one can choose the state together with $\varphi$ and solve for $\E$ (with the restriction that $\E>0$ at all points). In this way, there are many possible phase times. A choice of this variable can be seen as a choice of gauge.

\subsection{\label{sec:gf}Gauge independence and Faddeev--Popov gauge fixing}
From Eqs.~\eref{decomp-metric} to~ \eref{momentum-time}, as well as Eqs.~\eref{X-flow-q},~\eref{X-flow-p} and the identity
\eq{\label{eq:d-ds-X}
\del{}{s} = \E\tilde{g}^{\alpha\beta}\del{S}{x^{\alpha}}\del{}{x^{\beta}} = \E g^{\mu\nu}\del{S}{q^{\mu}}\del{}{q^{\nu}} = X \ ,
}
we find
\eq{\label{eq:X-flow-f-gen}
\del{f}{s} = Xf = \E\{f, H\}_{p = \partial S/\partial q} \ ,
}
for a function $f(q)\equiv \tilde{f}(q, p=\nabla S)$ and $H = g^{\mu\nu}p_{\mu}p_{\nu}/2+\V$. This equation is invariant under local reparametrizations of the internal time $s$, as it retains its functional form under the transformations $s\mapsto s'(s)$ and $\E\mapsto\E' = \E\D s/\D s'$. This is a manifestation of the time-reparametrization invariance of the theory, which constitutes a local (or gauge) symmetry because the time parameter can be changed locally without altering the form of the dynamical equations. This will also be clearly seen in the classical limit discussed in Sec.~\ref{sec:ehrenfest}. A gauge choice is thus made by fixing a particular choice of time. In the present formalism, this is achieved by fixing $\varphi$, thereby also fixing the coordinate $s$.

Let us rewrite the probabilities obtained from Eq. \eref{probs} as
\begin{align}\label{eq:probs-delta}
\!\!\D^d x\,p_{\Psi}(x) &= \D^{d+1}q\,\delta(\varphi(q)-s)\frac{\sqrt{|g|}R^2}{\E}\\
&= \D^{d+1}q\,\delta(\varphi(q)-s)\sqrt{|g|}R^2g^{\mu\nu}\del{S}{q^{\mu}}\del{\varphi}{q^{\nu}}\,,\notag
\end{align}
where we used Eq. \eref{E-phase-grad}. With this, the probabilities in Eq. \eref{probs-delta} do not depend explicitly on the configuration-space foliation but rather on the original configuration coordinates $q^{\mu}$ and the choice of $\varphi(q)$.\footnote{We take the amplitude $R$ to be a configuration-space scalar.}

To see that Eq.~\eref{probs-delta} is the analogue of the usual Faddeev--Popov gauge fixing in ordinary gauge theories, we first note that, at the level of expectation values (and also in the classical limit), the momenta can be identified with the components of the gradient of $S$, as we will see in Secs.~\ref{sec:expectations} and \ref{sec:ehrenfest}. With $p = \partial S/\partial q\equiv\nabla S$, a phase-space function $\tilde{\varphi}(q,p)$ can be used to define a choice of phase time via $\varphi(q) = \tilde{\varphi}(q, p = \nabla S)$. In this way, the fact that $s$ is an internal time (a time variable defined from the canonical or phase-space variables and not external to the system) corresponds to the fact that a complete gauge fixation of the local time-reparametrization symmetry can be obtained by what is called a ``canonical gauge condition,'' which is a gauge condition that necessarily involves the phase-space variables.\footnote{See, for example, \cite{HT:book} for a discussion of such gauge conditions.} The factor of $\delta(\varphi(q)-s)$ can thus be seen as a gauge-fixing Dirac delta distribution.

Second, this gauge-fixing distribution is accompanied by the factor $1/\E$, which, due to Eqs.~\eref{X-flow-q},~\eref{X-flow-p}, and~\eref{E-phase-grad}, is equivalent to $\{\tilde{\varphi},H\}_{p = \partial S/\partial q}$ with $H = g^{\mu\nu}p_{\mu}p_{\nu}/2+\V$. This is the Faddeev--Popov ``determinant.''\footnote{See \cite{HT:book} for a discussion of the Faddeev--Popov determinant in canonical gauge theories. In general, it can be defined as $\det\{\tilde{\varphi}_{a},G_b\}_{G = 0}$, where $\tilde{\varphi}_a$ are canonical gauge conditions and $G_b$ are gauge constraints. Here, we are dealing with only one local symmetry (time-reparametrization invariance with gauge constraint $H = 0$), and thus the Faddeev--Popov determinant simply coincides with $\{\tilde{\varphi},H\}_{H = 0}$.} The restriction $\E>0$ then corresponds to the restriction that this determinant must have a constant sign (which is chosen positive for convenience) so as to avoid the ``Gribov problem'' \cite{HT:book}. Indeed, the condition $\{\tilde{\varphi},H\}_{p = \partial S/\partial q}\neq0$ corresponds to the fact that the fixation of $\varphi$ is a complete gauge fixing; i.e., there exists no time reparametrization besides the identity transformation that preserves this choice, and thus $\varphi = s$ defines the time variable without ambiguity. Furthermore, an admissible gauge condition must not only be complete but also accessible: there must exist a well-defined time reparametrization (or, more precisely, an integral curve of $g^{\mu\nu}\partial_\mu S\partial_\nu$) that connects an arbitrary initial point $q^{\mu}$ to one that satisfies $\varphi(q)=s$. In this way, the choice of $\varphi$, and thus of $\E$, is not completely arbitrary, but it is subject to such admissibility conditions.

As in any gauge theory, we expect that the physical predictions of our formalism do not depend on a choice of gauge. As the choice of $\varphi$ fixes the gauge, we can verify that the probabilities do not change if we vary $\varphi$. For any region $\mathscr{R}$ of the configuration space, we obtain
\begin{align}
&\delta\int_{\mathscr{R}}\D^{d+1}q\,\delta(\varphi(q)-s)\sqrt{|g|}R^2g^{\mu\nu}\del{S}{q^{\mu}}\del{\varphi}{q^{\nu}}\notag\\
&=\int_{\mathscr{R}}\D^{d+1}q\,\delta\varphi\delta'(\varphi(q)-s)\sqrt{|g|}R^2g^{\mu\nu}\del{S}{q^{\mu}}\del{\varphi}{q^{\nu}}\notag\\
&\ \ +\int_{\mathscr{R}}\D^{d+1}q\,\delta(\varphi(q)-s)\sqrt{|g|}R^2g^{\mu\nu}\del{S}{q^{\mu}}\del{\delta\varphi}{q^{\nu}}\notag\\
&=\int_{\mathscr{R}}\D^{d+1}q\,\delta\varphi\delta'(\varphi(q)-s)\sqrt{|g|}R^2g^{\mu\nu}\del{S}{q^{\mu}}\del{\varphi}{q^{\nu}}\notag\\
&\ -\int_{\mathscr{R}}\D^{d+1}q\,\delta\varphi\del{}{q^{\nu}}\left[\delta(\varphi(q)-s)\sqrt{|g|}R^2g^{\mu\nu}\del{S}{q^{\mu}}\right]\notag\\
&=\int_{\mathscr{R}}\D^{d+1}q\,\delta\varphi\sqrt{|g|}R^2g^{\mu\nu}\del{S}{q^{\mu}}\del{\varphi}{q^{\nu}}\notag\\
&\ \ \ \ \times\left[\delta'(\varphi(q)-s)-\delta'(\varphi(q)-s)\right]\notag\\
&=0 \ ,
\end{align}
where the prime indicates the derivative of the Dirac delta distribution with respect to its argument, and we used the conservation of the Klein--Gordon current $j^{\mu}[\Psi]$ [cf. Eq. \eref{polar-KG2}] to arrive at the next-to-last line. In this way, the probabilities are invariant under changes of gauge, and thus they are independent of the choice of $\varphi$ (at least considering transformations of $\varphi$ that are continuously connected).

Furthermore, the leaf coordinates $x^i$ are also gauge independent. In the $(s,x)$ coordinate system, the condition $\partial x^i/\partial s = 0$ translates to
\eq{\label{eq:x-inv}
\E g^{\mu\nu}\del{S}{q^{\mu}}\del{x^i}{q^{\nu}} = 0
}
in the original coordinates. As any admissible $\E > 0$ can be factored out of this equation, the configuration-space functions $x^i(q)$ obey the same equation regardless of the choice of $\varphi$, and thus do not vary with the choice of gauge. For this reason, the $s$ independence of the probability density given in Eq. \eref{probs} also indicates that the probabilities $p_{\Psi}(x)\D^dx$ refer to gauge-independent quantities.

\subsection{\label{sec:weyl}Conformal (Weyl) invariance}
If the potential in Eq.~\eref{KG} is of the form
\eq{\label{eq:potential-conformal}
V(q) &= V_{\rm cl}(q)-\frac{\hbar^2 [(d+1)-2]}{8[(d+1)-1]}\mathcal{R}\\
&= V_{\rm cl}(q)-\frac{\hbar^2 (d-1)}{8d}\mathcal{R} \ ,
}
where $V_{\rm cl}$ is some classical potential, and $\mathcal{R}$ is the Ricci scalar of configuration space, which we recall has $d+1$ dimensions (the leaves $\Sigma$ have $d$ dimensions), then it is possible to show that Eq.~\eref{KG} is covariant under conformal (Weyl) transformations of the configuration-space metric \cite{Kiefer:book,Misner,Halliwell}, $g_{\mu\nu}\mapsto \Omega^2g_{\mu\nu}$,\footnote{Because of the Laplace--Beltrami factor ordering, the quantum constraint in Eq.~\eref{KG} is already form invariant under general transformations of the configuration-space coordinates, including those that may rescale the metric by a factor. The potential $V(q)$ does not change under these transformations, as it is a configuration-space scalar ($\varphi$ and $S$ are also taken to be scalars). In contrast, the Weyl transformation considered here corresponds to an independent rescaling of the metric (not compensated by a coordinate transformation).} provided the potential and the wave function are also transformed by
\eq{
V(q) &\mapsto \Omega^{-2}V(q) \ , \\
\Psi &\mapsto \Omega^{-\frac{[(d+1)-2]}{2}}\Psi = \left[\Omega^{-\frac{(d-1)}{2}}R\right]\e^{\frac\I\hbar S} \ .
}
As $\Omega$ is taken to be real and positive, this corresponds to a rescaling of the amplitude $R$ of the quantum state. It is reasonable to adopt the factor ordering that leads to the potential in Eq.~\eref{potential-conformal}, as this conformal symmetry is also present at the classical level as a consequence of local time-reparametrization invariance: it corresponds to a rescaling of the $\E$ field [as can be seen from the forthcoming Eqs.~\eref{Ehren-rel1c} and \eref{Ehren-rel2c} -- see also Eqs.~\eref{X-flow-q}, \eref{X-flow-p}, and \eref{X-flow-f-gen}].

In this case, we find that the probabilities in Eq.~\eref{probs-delta} transform as
\eq{
\D^d x\,p_{\Psi}(x)&\mapsto \D^{d+1}q\,\delta(\varphi(q)-s)\left(\Omega^{d+1}\sqrt{|g|}\right)\\
&\ \ \ \times\left(\Omega^{-d+1}R^2\right)\left(\Omega^{-2}g^{\mu\nu}\right)\del{S}{q^{\mu}}\del{\varphi}{q^{\nu}}\\
& = \D^d x\,p_{\Psi}(x) \ .
}
In other words, the factors of $\Omega$ are canceled, and the probabilities are invariant under these conformal transformations. This is to be expected, as the probabilities are also gauge independent (in the sense of Sec.~\ref{sec:gf}).

We now turn to the definition of the basic expectation values and subsequently discuss their Ehrenfest dynamics.

\subsection{\label{sec:expectations}Expectation values}
The expectation value of a configuration-space function $f(q(s,x))$ is given by
\eq{\label{eq:av}
\braket{f(s)} &:= \int_{\Sigma}\D^{d}x\ p_{\Psi}(x)f(q(s,x))\\
&=\int_{\Sigma}\D^{d}x\ \frac{N\sqrt{h}}{\E}\, \left.\Psi^* f\Psi\right|_{q(s,x)}\ ,
}
and this leads to the average of the configuration point $q^{\mu}(s,x)$ as a particular case. Notice that this has the usual form of the expectation value of a configuration-space operator. 

Similarly, the averages of the momenta conjugate to the $x^i$ variables can be obtained if we set $f = \partial S/\partial x^{i}$ because
\eq{\label{eq:av-momentum}
\braket{P_{i}(s)} &= \int_{\Sigma}\D^{d}x\ \frac{N\sqrt{h}}{\E}\, \Psi^* \del{S}{x^{i}}\Psi\\
 &= \int_{\Sigma}\D^{d}x\ \frac{N\sqrt{h}R^2}{\E}\,\del{S}{x^{i}}\\
 &\ \ \ +\frac{\hbar}{2\I}\int_{\Sigma}\D^{d}x\ \del{}{x^{i}}\left(\frac{N\sqrt{h}R^2}{\E}\right)\\
 &= \int_{\Sigma}\D^{d}x\ \frac{N\sqrt{h}}{\E}\, \Psi^*\,\hat{P}_{i}\Psi \ ,
}
where the momentum operator is defined in analogy to the ``nonrelativistic'' definition in Eq. \eref{momentum} \cite{DeWitt57}:
\eq{\label{eq:momentum-op}
\hat{P}_{i} = \frac\hbar\I \left(\frac{N\sqrt{h}}{\E}\right)^{-\frac12}\del{}{x^{i}}\left(\frac{N\sqrt{h}}{\E}\right)^{\frac12} \ ,
}
and we used the assumption that the square-integrable amplitude leads to a vanishing integral of $\frac{\hbar}{2\I}\del{}{x^i}\left(N\sqrt{h}R^2/\E\right)$. Analogously, we can use $0 = \frac{\hbar}{2\I}\del{}{s}\left(N\sqrt{h}R^2/\E\right)$ [cf. Eq. \eref{conserve-integrand}] to establish
\eq{\label{eq:av-momentum-0}
\braket{P_0(s)} &:= \int_{\Sigma}\D^{d}x\ \frac{N\sqrt{h}}{\E}\, \Psi^* \del{S}{s}\Psi\\
 &= \int_{\Sigma}\D^{d}x\ \frac{N\sqrt{h}}{\E}\, \Psi^*\,\hat{P}_{0}\Psi \ ,
}
with $\hat{P}_0$ defined similarly to Eq. \eref{momentum-op}.

Furthermore, for the momenta conjugate to the original coordinates $q^{\mu}$, if instead of the ``nonrelativistic'' definition in Eqs. \eref{momentum} and \eref{momentum-polar} we use the definition \cite{DeWitt57}
\eq{\label{eq:momentum-change-coords}
\hat{p}_{\mu} := \frac12\left[\hat{P}_{\nu},\del{x^{\nu}}{q^{\mu}}\right]_++\frac{\I\hbar}{2}\del{}{s}\left(\del{s}{q^{\mu}}\right) \ ,
}
where $[\cdot,\cdot]_+$ is the anticommutator, we obtain
\eq{\notag
\!\!\!\frac{\hat{p}_{\mu}\Psi}{\Psi} \!=\! \del{S}{q^{\mu}} +\frac{\hbar}{2\I}\left[\del{}{q^{\mu}}\log\frac{N\sqrt{h}R^2}{\E}+\del{}{x^{i}}\left(\del{x^{i}}{q^{\mu}}\right)\right] \,,
}
which leads to
\eq{\label{eq:av-momentum-mu}
\braket{p_{\mu}(s)} &= \int_{\Sigma}\D^{d}x\ \frac{N\sqrt{h}}{\E}\, \Psi^* \del{S}{q^{\mu}}\Psi\\
 &= \int_{\Sigma}\D^{d}x\ \frac{N\sqrt{h}}{\E}\, \Psi^*\,\hat{p}_{\mu}\Psi \ ,
}
because of Eq. \eref{conserve-integrand} and the assumption that the integral of $\frac{\hbar}{2\I}\del{}{x^i}\left(\del{x^i}{q^{\mu}}N\sqrt{h}R^2/\E\right)$ vanishes. In this way, the averages of momenta in Eqs. \eref{av-momentum}, \eref{av-momentum-0}, and \eref{av-momentum-mu} have the familiar form of expectation values of momentum operators.\footnote{The expectation values given in Eqs.~\eref{av} and~\eref{av-momentum-mu} are particular cases of averages of differential operators with respect to the $x^i$ variables, $\braket{O(s)}:= \int_{\Sigma}\D^dx'\int_{\Sigma}\D^dx \,(N\sqrt{h}\Psi^*/\E)_{s,x'}\hat{O}(s;x',x)(N\sqrt{h}\Psi/\E)_{s,x}$. In the case in which the operator is self-adjoint with respect to the integration measure on the $x^i$ variables, one can use its spectral decomposition to write $\braket{O(s)} = \sum_\ell \ell \,p_{\Psi}(\ell)$, where $\ell$ are the eigenvalues of $\hat{O}$ (one should integrate rather than sum over them if they are continuous) and $p_{\Psi}(\ell;s) := |\int_{\Sigma}\D^dx\,(N\sqrt{h}\psi_{\ell}^*\Psi/\E)_{s,x}|^2$ could be interpreted as the probability density to observe the value $\ell$ at the instant $s$. It is defined in terms of the eigenfunctions $\psi_{\ell}(s,x)$ of $\hat{O}$, which form a complete orthonormal system with respect to the integration measure over the $x^i$ variables; i.e., they satisfy $\sum_\ell \psi_{\ell}(s,x')\psi_{\ell}^*(s,x) = \delta^d(x',x)\E/(N\sqrt{h})$ and $\int_{\Sigma}\D^dx\,N\sqrt{h}\psi_{\ell'}^*(s,x)\psi_{\ell}(s,x)/\E = \delta(\ell',\ell)$. Although these eigenfunctions need not solve the constraint given in Eq.~\eref{KG}, if one has a set of solutions $\Psi_\alpha$ to Eq.~\eref{KG} that is complete and orthonormal with respect to the Klein--Gordon inner product defined in Eq.~\eref{KGprod}, one can write the solution $\Psi$ as the formal superposition $\Psi(q) = \sum_\alpha \xi_\alpha\Psi_\alpha(q)$. Given the foliation induced by $\Psi$ [cf. Sec.~\ref{sec:fol}] and, in particular, given $\E$, we can also define another solution to Eq.~\eref{KG} with $\Psi_\ell(q) = \sum_\alpha \xi_{\ell,\alpha}\Psi_\alpha(q)$, where $\xi_{\ell,\alpha}:= \int_{\Sigma}\D^dx\,(N\sqrt{h}\psi_{\ell}\Psi^*_\alpha/\E)_{s,x}$. In this way, we can interpret $|(\Psi_{\ell},\Psi)|^2 = |\sum_\alpha \xi^{*}_{\ell,\alpha}\xi_\alpha|^2 = |\int_{\Sigma}\D^dx\,(N\sqrt{h}\psi_{\ell}^*\Psi/\E)_{s,x}|^2 = p_{\Psi}(\ell;s)$ as the transition probability between two physical states. The rigorous definition of transition probabilities, however, requires a detailed examination of an appropriate notion of state updates relative to measurements defined at a given moment of the internal time (and the explicit construction of a physical ``basis'' $\Psi_{\alpha}$ would also be of interest). See the comments in Sec.~\ref{sec:conclusions}.}

\subsection{\label{sec:ehrenfest}The Ehrenfest theorem}
Do the expectation values of the configuration point $q^{\mu}$ and the momenta satisfy a type of Ehrenfest theorem? First, notice that the gauge independence of the probabilities $\D^d x\,p_{\Psi}(x)$ defined in Eq. \eref{probs} implies that the integration measure only involves the $x^i$ variables, and no integration over the internal time $s$ is performed. For this reason, the Hamiltonian $\hat{H}$ [which includes a Laplace--Beltrami operator that involves all configuration-space variables, cf. Eq. \eref{KG}] is no longer self-adjoint with respect to the integration measure used to compute expectation values, in contrast to the nonrelativistic case reviewed in Sec. \ref{sec:nonrel}. As a result, the analogue of the Ehrenfest theorem in this context is not readily written in terms of a commutator of operators with a self-adjoint Hamiltonian $\hat{H}$ in the same form as in Eqs. \eref{Ehrencom1} and \eref{Ehrencom2} in the nonrelativistic case. Nevertheless, we will see that it is not only possible to examine the dynamics of expectation values, but also that the correct classical limit can be obtained.

From Eqs.~\eref{conserve-integrand} and~\eref{av}, we obtain
\eq{\label{eq:av1}
\del{}{s}\braket{f(s)} =\int_{\Sigma}\D^{d}x\ \frac{N\sqrt{h}}{\E}R^2\, \del{f}{s} \ .
}
Because of the identity in Eq.~\eref{d-ds-X}, we find
\eq{\label{eq:Ehren-rel1}
\del{}{s}\braket{f(s)} &=\int_{\Sigma}\D^{d}x\ \sqrt{|\tilde{g}|}R^2\, g^{\mu\nu}\del{S}{q^{\mu}}\del{f}{q^{\nu}} \ ,
}
which is the direct analogue of the nonrelativistic equation \eref{Ehren1}. As in the nonrelativistic case, if $\Psi(q(s,x))$ is a narrowly peaked wave packet at an instant $s$ such that $N\sqrt{h}R^2/\E$ approximates a Dirac delta distribution centered at $x = x_0$, then the Ehrenfest equation \eref{Ehren-rel1} leads to the counterpart to Eq. \eref{Ehren1c}:
\begin{align}
\del{}{s}\braket{f(s)} &\simeq \left(\E g^{\mu\nu}\del{S}{q^{\mu}}\del{f}{q^{\nu}}\right)_{q = q(s,x_0)}\label{eq:Ehren-rel1c}\\
&= \{f,\E H\}_{q = q(s,x_0),p = \partial S/\partial q|_{q = q(s,x_0)}}\ ,\notag
\end{align}
at the instant $s$, where the momenta are again given by the gradient of $S$, and the Hamiltonian phase-space function is $\E H = \E(g^{\mu\nu}p_{\mu}p_{\nu}/2+\V)$, which vanishes when evaluated at $p = \partial S/\partial q \equiv \nabla S$ due to Eq.~\eref{polar-KG1}.\footnote{Notice that $\{\cdot,\E H\}_{H = 0} = \E_{H = 0}\{\cdot,H\}_{H = 0}$.} As discussed in Sec.~\ref{sec:gf}, the functions $\varphi$ and $\E$ may coincide with general phase-space functions via $\varphi(q)= \tilde{\varphi}(q,p=\nabla S)$ and $\E(q) =  \tilde{\E}(q,p=\nabla S) = 1/\{\tilde{\varphi},H\}_{p = \partial S/\partial q}$. The function $\E$ corresponds to a choice of einbein on the relativistic particle's worldline or a choice of lapse function in minisuperspace models of general relativity. The Ehrenfest equation of motion for the configuration point $q^{\mu}$ is obtained from Eq. \eref{Ehren-rel1c} by setting $f = q^{\mu}(s,x)$ for each value of $\mu$, and, as before, a classical limit that is independent of $\hbar$ can be obtained if $S$ and $R$ admit a series expansion in powers of $\hbar$ with lowest order $\hbar^0$. 

The Ehrenfest dynamics of momenta can be obtained analogously as follows. From Eqs. \eref{conserve-integrand}, \eref{d-ds-X}, and \eref{av-momentum-mu}, we find
\eq{
\del{}{s}\braket{p_{\mu}(s)} &= \int_{\Sigma}\D^{d}x\ N\sqrt{h}R^2\, g^{\rho\sigma}\del{S}{q^{\rho}}\del{^2S}{q^{\sigma}\partial q^{\mu}} \ ,
}
which, because of Eq.~\eref{p-X-id}, leads to the direct counterpart to Eq. \eref{Ehren2}:
\eq{\label{eq:Ehren-rel2}
&\del{}{s}\braket{p_{\mu}(s)}\\
&= -\int_{\Sigma}\D^{d}x\ \sqrt{|\tilde{g}|}R^2\, \left(\frac12\del{g^{\rho\sigma}}{q^{\mu}}\del{S}{q^{\rho}}\del{S}{q^{\sigma}}+\del{\V}{q^{\mu}}\right) \ ,
}
except now the integration is performed over the $x^i$ variables on the leaf $\Sigma$. Nonetheless, the correct classical limit may be obtained if, again, the wave function at the instant $s$ is narrowly peaked, so as to yield $N\sqrt{h}R^2/\E$ as an approximation of the Dirac delta distribution centered at $x = x_0$. Then, the Ehrenfest equation \eref{Ehren-rel2} leads to the counterpart of Eq. \eref{Ehren2c}:
\begin{align}\notag
\del{}{s}\braket{p_{\mu}(s)}&\simeq -\left[\E\left(\frac12\del{g^{\rho\sigma}}{q^{\mu}}\del{S}{q^{\rho}}\del{S}{q^{\sigma}}+\del{\V}{q^{\mu}}\right)\right]_{q = q(s,x_0)}\\ \label{eq:Ehren-rel2c}
& = \{p_{\mu},\E H\}_{q = q(s,x_0),p=\partial S/\partial q|_{q = q(s,x_0)}} \,.
\end{align}
If an expansion in powers of $\hbar$ is possible, we once again recover classical equations and classical solutions that are independent of $\hbar$. Thus, a classical solution emerges from the dynamics of expectation values given by Eqs. \eref{Ehren-rel1c} and \eref{Ehren-rel2c}, at least when the probability distribution defined by Eq. \eref{probs} is narrowly peaked. Notice that the equations of motion obeyed by this solution exhibit a local time-reparametrization invariance: one can freely redefine the $s$ parameter as long as $\E$ is correspondingly adjusted (cf. Sec. \ref{sec:gf}). The equations of motion remain of the same form. Finally, we see from Eqs.~\eref{Ehren-rel1c} and \eref{Ehren-rel2c} that the classical limit of expectation values leads to the classical trajectory as one particular, approximate solution to Eqs.~\eref{X-flow-q} and \eref{X-flow-p} that is singled out by a narrowly peaked probability distribution. 

\subsection{\label{sec:GP}Relation to the previous phase-time construction}
We now discuss how the previous construction due to Padmanabhan and Greensite, which was presented in \cite{Padmanabhan,Greensite2} (see also  \cite{Greensite,Greensite1}), is recovered as a particular case of the formalism presented here, and we also comment on the difficulties found in their approach.\footnote{The reader is also referred to the review and critique of this approach by Brotz and Kiefer \cite{Brotz}.} To begin with, Greensite and Padmanabhan only consider the choice $\E = 1$, and so are restricted to internal time functions $\varphi(q)$ that satisfy $g^{\mu\nu}\partial_{\mu}S\partial_{\nu}\varphi = 1$ [cf. Eq.~\eref{E-phase-grad}]. Moreover, they assume that the $x^i$ coordinates can be chosen so as to yield $\tilde{g}^{0i} = 0$. From Eqs.~\eref{decomp-inv-metric} and~\eref{S-N-beta}, together with the regularity of $h_{ij}$, we see that this corresponds to assuming that it is possible to define these coordinates so that $\partial S/\partial x^i = 0$. This means that $S$ is only a function of $s$, and, due to Eqs.~\eref{S-N-beta} and~\eref{momentum-time} with $\E = 1$, the same holds for $N$ and $\V$, both of which must be positive in this case. Using Eqs.~\eref{decomp-metric},~\eref{decomp-inv-metric},~\eref{S-N-beta}, and~\eref{momentum-time} we also find
\eq{\label{eq:GP1}
\tilde{g}^{00}\del{S}{s} = 1 \ ,
}
which was a key condition in the the Greensite--Padmanabhan approach. With these coordinates, the Laplace--Beltrami operator can be decomposed into
\eq{\label{eq:nabla2GP}
\nabla^2 &= \mathrm{D}_0^2+\mathrm{D}^2 \ ,\\
\mathrm{D}_0^2 &:= -\frac{1}{N\sqrt{h}}\del{}{s}\left(\frac{\sqrt{h}}{N}\del{}{s}\right)\ ,\\
\mathrm{D}^2 &:= \frac{1}{\sqrt{h}}\del{}{x^i}\left(\sqrt{h}h^{ij}\del{}{x^j}\right) \ .
}
This decomposition was used to construct a type of ``phase-time Sch\"odinger equation,'' which relates the $s$ derivative of the wave function $\Psi$ to an effective Hamiltonian (which is generally dependent on $\Psi$). Indeed, if we use Eqs.~\eref{momentum-time},~\eref{nabla2S}, and~\eref{conserve-integrand} with the Greensite--Padmanabhan assumptions, we find
\eq{\label{eq:schro-GP}
\I\del{\Psi}{s} &= \left(\frac{\I}{R}\del{R}{s}-\del{S}{s}\right)\Psi\\
&= \left[-\frac{\I}{2N\sqrt{h}}\del{}{s}\left(N\sqrt{h}\right)-\del{S}{s}\right]\Psi\\
&= \left[-\frac{1}{2}\mathrm{D}^2+\tilde{\V}-\frac{\I}{2N\sqrt{h}}\del{}{s}\left(N\sqrt{h}\right)\right]\Psi \ ,
}
where we have set $\hbar = 1$ for simplicity, and we defined
\eq{
\tilde{\V}&:= \frac{1}{2\Psi}\mathrm{D}^2\Psi-\del{S}{s}\\
&= \frac{1}{\Psi^*\Psi}\left[\frac{1}{2}\mathrm{Re}\left(\Psi^*\mathrm{D}^2\Psi\right)-\mathrm{Im}\left(\Psi^*\del{\Psi}{s}\right)\right]\ ,
}
with the last equality following from $\partial S/\partial x^i = 0$. The last line of Eq.~\eref{schro-GP} is exactly the effective Schr\"odinger equation obtained by Greensite and Padmanabhan with a different method. Notice its formal similarity to Eq. \eref{nonrel-H} in the nonrelativistic case. Because of Eqs.~\eref{S-N-beta} and ~\eref{momentum-time}, the ``measure factor'' in the effective Hamiltonian can also be written as $-\I/2$ times:
\begin{align}
&\del{}{s}\log\left(N\sqrt{h}\right)\notag\\
&= -\frac{1}{N\sqrt{h}R^2}\del{}{s}\left(\frac{\sqrt{h}}{N}R^2\del{S}{s}\right)-\frac{1}{R^2}\del{R^2}{s}\notag\\
&= \frac{1}{\Psi^*\Psi}\left[-\frac{1}{N\sqrt{h}}\del{}{s}\left(\frac{\sqrt{h}}{N}\mathrm{Im}\Psi^*\del{\Psi}{s}\right)-\del{}{s}\left(\Psi^*\Psi\right)\right]\notag\\
&= \frac{1}{\Psi^*\Psi}\left[\mathrm{Im}\left(\Psi^*\mathrm{D}_0^2\Psi\right)-\del{}{s}\left(\Psi^*\Psi\right)\right] \ .
\end{align}
The last line of this equation was the expression that was adopted in their construction for the measure factor. However, because of Eq.~\eref{conserve-integrand}, the first factor in this last line is superfluous as it is equal to zero.

With the effective phase-time Schr\"odinger equation, the goal was to prove the Ehrenfest equations for the coordinates $x^i$ and their momenta $\hat{P}_i$. For $x^i$, they proved what was referred to as the ``first Ehrenfest equation,'' which expressed the $s$ derivative of the expectation value of $x^i$ in terms of the expectation value of the commutator of $x^i$ with the constraint operator $\hat{H}$. Here, we can prove this by using Eq.~\eref{identity-com} instead of resorting to the effective Schr\"odinger equation~\eref{schro-GP}. Indeed, we can use the results of Sec.~\ref{sec:fol} together with Eq.~\eref{identity-com} and the Greensite--Padmanabhan assumptions listed above to obtain the identity
\eq{\label{eq:identity-com-GP}
R^2\,\del{x^i}{s} &=\Psi^*\I[\hat{H},x^i]\Psi +\frac{\I}{2\sqrt{h}}\del{}{x^{j}}\left(\sqrt{h}R^2\,h^{ij}\right)\,,
}
which then leads to
\eq{\label{eq:first-Ehrenfest}
\del{}{s}\braket{x^i} = \int_{\Sigma}\D^dx\,N\sqrt{h}\,\Psi^*\I [\hat{H},x^i]\Psi\ ,
}
because of Eq.~\eref{conserve-integrand} and the fact that $N$ does not depend on $x^i$ on account of the Greensite--Padmanabhan assumptions. The equation~\eref{first-Ehrenfest} is the first Ehrenfest equation of Greensite and Padmanabhan. However, one can readily see that this equation is trivial: since $\partial x^i/\partial s = 0$, the left-hand side of Eqs.~\eref{identity-com-GP} and~\eref{first-Ehrenfest} is zero. Likewise, because of the relation [cf. Eqs.~\eref{com} and~\eref{identity-com-GP} and the Greensite--Padmanabhan assumptions]
\eq{
\Psi^*[\hat{H},x^i]\Psi = -\frac{1}{2\sqrt{h}}\del{}{x^j}\left(\sqrt{h}R^2\,h^{ij}\right) \ ,
}
the right-hand side of Eq.~\eref{first-Ehrenfest} integrates to a boundary term, which can be assumed to vanish. In this way, there is no dynamical content in that equation.\footnote{This was also observed by Brotz and Kiefer \cite{Brotz} in another way, and they remarked that $x^i$ could be considered ``perennials'' (following a terminology adopted by Kucha\v{r}; see, e.g., \cite{Kuchar}). As we have seen in Sec.~\ref{sec:gf}, the $x^i$ variables are indeed gauge independent (in the sense discussed there).}

The ``second Ehrenfest equation,'' which concerns the dynamics of $\braket{P_i}$, was more difficult to obtain in their approach. Rather than a simple application of the effective Schr\"odinger equation~\eref{schro-GP}, it required extra assumptions, such as requiring that the phase has a ``rapid variation in the $s$ direction,'' or, as Brotz and Kiefer discussed \cite{Brotz}, the condition
\eq{
\left<\del{^2}{x^i\partial s}\log\sqrt{|\tilde{g}|}\right> = 0
}
should be imposed. Rather than examining the imposition of these extra requirements, we note that the expectation values of $\hat{P}_i$ are also devoid of dynamics under the Greensite--Padmanabhan assumptions. Indeed, due to $\tilde{g}^{0i} = 0$, which implies $\partial S/\partial x^i = 0$ as we have seen, we find from Eq.~\eref{av-momentum} that $\braket{P_i} = 0$ for all $s$. In this way, both the first and second Ehrenfest equations are trivial in this approach. In contrast, the formalism presented here not only generalizes this previous construction, but it is also able to capture the dynamics of the expectation values of the original variables ($q^{\mu}$ and $p_{\mu}$) and their classical limit [cf. Eqs.~\eref{Ehren-rel1c} and~\eref{Ehren-rel2c}].

\section{\label{sec:qg}Canonical quantum gravity}
The formalism discussed in Sec.~\ref{sec:rel} is well suited for mechanical models with local time-reparametrization invariance. Field-theoretic models, such as canonical quantum gravity, can be formally treated in the same way: one can use the phase of the wave functional to define a foliation of the field space and a conserved, gauge-independent inner product as the one given in Eq.~\eref{fol-norm}. More precisely, the wave functional in canonical quantum gravity must be invariant under three-dimensional diffeomorphisms (on spacelike leaves\footnote{The spacetime foliation of canonical gravity is not to be confused with the field (configuration) space foliation involved in the definition of the probabilities in Eqs.~\eref{probs} and~\eref{probs-qg} (see Sec.~\ref{sec:fol}).}) and it must solve the local Wheeler--DeWitt constraint\footnote{The presence of a general matter Hamiltonian in Eq.~\eref{WDW} can be treated by means of the Born--Oppenheimer technique \cite{LapRuba,Banks,Brout1,Zeh,Venturi,KieferSingh,KieferRev,Chataignier:2019,Chataignier:2022,Chataignier:2023}. Here, we disregard this for simplicity. Moreover, we have restored factors of $\hbar$.}:
\eq{\label{eq:WDW}
0 = \hat{\mathscr{H}}\Psi = -\frac{\hbar^2}{2}\,{\text{\Huge``}}G_{abcd}\frac{\delta^2\Psi}{\delta \mathtt{h}_{ab}\delta \mathtt{h}_{cd}}{\text{\Huge''}}\,+V(\mathtt{h})\Psi \ ,
}
which is the field-theoretic analogue of Eq.~\eref{KG}. In Eq.~\eref{WDW}, $G_{abcd}$ are the coefficients of an inverse metric on the field-configuration space,\footnote{The exact expression for the configuration-space metric is well known (see \cite{Kiefer:book,DeWitt}). For convenience, we have absorbed factors of Newton's constant $G$ and the speed of light $c$ in $G_{abcd}$ and in $V(\mathtt{h})$.} whereas $\mathtt{h}_{ab}$ are the coefficients of the induced metric on the spatial leaves (not to be confused with the configuration-space metric coefficients $h_{ij}$ discussed in Sec.~\ref{sec:fol}). The quotation marks in the kinetic term indicate that we must choose a factor ordering as well as a procedure to regularize the second functional derivative, which is evaluated at a single spacetime point. The task of defining a consistent and satisfactory regularization of the Wheeler--DeWitt constraints is an active topic of research for which a consensus has not yet been reached (see, e.g., \cite{Tsamis:1987,Williams:1997,Ambjorn:1998,Hamber:2011,Liu:2016,Feng:2018,Ambjorn:2022,Lang:2023a,Lang:2023b,Lang:2023c} and also \cite{Kiefer:book} for further details and references on this point).

Within a formal treatment (i.e., assuming that a consistent regularization is possible), a direct application of the formalism of Sec.~\ref{sec:rel} leads to the definition of the squared norm
\eq{\label{eq:fol-norm-qg}
(\Psi,\Psi) = \int_{\Sigma}\mathscr{D}x\ \frac{N\sqrt{h} R^2}{\E} \ ,
}
where the integral of Eq.~\eref{fol-norm} is now replaced by a functional integration, and the $x^i$ variables are now fields that are gauge independent in the sense that they obey the field-theoretic version of Eq.~\eref{x-inv} \cite{Kuchar,Isham,Chataignier:2023}:
\eq{\label{eq:x-inv-qg}
G_{abcd}\frac{\delta S}{\delta \mathtt{h}_{ab}}\frac{\delta x}{\delta \mathtt{h}_{cd}} = 0 \ .
}
The squared norm in Eq.~\eref{fol-norm-qg} leads to the probability distribution
\eq{\label{eq:probs-qg}
p_{\Psi} := \frac{N\sqrt{h} R^2}{\E}
}
for normalized states. It is to be functionally integrated over the $x$ fields defined by Eq.~\eref{x-inv-qg}.

Ultimately, the formalism of Padmanabhan and Greensite was also intended to be applied to canonical quantum gravity. However, besides the difficulties that were encountered in that approach, no such formalism can produce reliable predictions until the factor ordering problems in Eq.~\eref{WDW} have been regularized \cite{Tsamis:1987}. Meanwhile, it is preferable to focus on toy models that can be more rigorously analyzed, such as the quantum cosmological models of homogeneous universes. In this setting, the problem of time can be very clearly analyzed (and solved), and tentative conclusions can be extrapolated to the full theory (which is still under construction). In what follows, we consider one such model as an example of the general formalism discussed in Sec.~\ref{sec:rel}.

\section{\label{sec:eg}A cosmological model}
To illustrate the above formalism, let us consider a simple cosmological model, where the spacetime metric is that of a Friedmann--Lama\^itre--Robertson--Walker universe with scale factor $a(\tau)$ and vanishing cosmological constant, and the matter content consists of a minimally coupled homogeneous scalar field $\phi(\tau)$. The Hamiltonian constraint for this model can be shown to be \cite{Kiefer:book}
\eq{
-\frac{2\pi G}{3V_0}\frac{p_a^2}{a}+\frac{p_{\phi}^2}{2a^3V_0}-\frac{3V_0}{8\pi G} k a = 0 \ ,
}
where we use units in which $c = 1$, the scalar field is taken to be massless and without self-interaction for simplicity, $p_a$ and $p_\phi$ are, respectively, the canonical momenta conjugate to $a$ and $\phi$, $G$ is Newton's constant, $k$ is the curvature parameter, and $V_0$ is a fiducial spatial volume that appears because of the spatial integration over homogeneous fields in the action. To simplify the notation, we can make the redefinitions (see, e.g., \cite{KTV})
\eq{
a\to \frac{a}{V_0^\frac13} \ , \ p_a\to p_a V_0^{\frac13} \ ,\ k\to \frac{k}{V_0^\frac23} \ , \ \kappa = \frac{4\pi G}{3} \ , 
}
to obtain the constraint
\eq{\label{eq:cosm-constraint}
-\frac{\kappa a^2}{2}p_a^2+\frac{p_{\phi}^2}{2}-\frac{k a^4}{2\kappa} = 0 \ ,
}
where now the scale factor has units of length and $k$ is dimensionless, whereas $\phi^2$ and $1/\kappa$ have units of mass/length, with the constraint in Eq.~\eref{cosm-constraint} having units of mass$\times$length$^3$. In the context of the previous phase-time analyses, such a constraint was also considered in \cite{Brotz,Greensite3}. Its quantization leads to
\eq{\label{eq:WDW-osc}
\frac{\kappa\hbar^2 a}{2}\del{}{a}\left(a\del{\Psi}{a}\right)-\frac{\hbar^2}{2}\del{^2\Psi}{\phi^2}-\frac{ka^4}{2\kappa}\Psi = 0 \ , 
}
with the Laplace--Beltrami ordering. Let us now analyze this constraint in two different cases.

\subsection{\label{sec:eg-part}Free massless relativistic particle}
If $k=0$, the constraint becomes that of a free massless relativistic particle in two ``spacetime'' dimensions, which is possibly the simplest instance of a time-reparametrization invariant system. Indeed, via the standard transformation (see, e.g., \cite{Kiefer:book}) $a = a_0\e^{\alpha}$, we obtain
\eq{\label{eq:WDW-part}
\frac{\hbar^2}{2}\del{^2\Psi}{\alpha^2}-\frac{\hbar^2}{2\kappa}\del{^2\Psi}{\phi^2} = 0 \ ,
}
where $\alpha$ is dimensionless and takes values over the real line.\footnote{In this way, the two-dimensional Minkowski ``spacetime'' corresponds to configuration space, where the ``time'' coordinate corresponds to $\alpha$ and the ``space'' coordinate is $\sqrt{\kappa}\phi$. Notice, however, that both coordinates are dimensionless. The constraint in Eq.~\eref{WDW-part} has units of mass$^2\times$length$^2$ with $c=1$.} As an example, consider then the solution
\eq{\label{eq:wf-part}
\Psi = \exp\left[\frac\I\hbar \mathtt{K}\left(\alpha+\sqrt{\kappa}\phi\right)-\frac{\left(\alpha+\sqrt{\kappa}\phi-x_0\right)^2}{2\epsilon^2}\right] \ ,
}
where $\mathtt{K},x_0,\epsilon$ are real parameters, and $\mathtt{K}$ has units of mass$\times$length, whereas $x_0,\epsilon$ are dimensionless. We also assume that $\mathtt{K}\neq0$ and $\epsilon>0$. The gauge-independent variable $x$ satisfies the equation $\partial x/\partial(\sqrt{\kappa}\phi)-\partial x/\partial\alpha = 0$ [cf. Eq.~\eref{x-inv}], and we can choose the solution 
\eq{
x = \alpha+\sqrt{\kappa}\phi \ ,
}
which is dimensionless and takes values over the real line. Likewise, we can define the internal time as a solution to the equation [cf. Eq.~\eref{E-phase-grad}]
\eq{\label{eq:internal-time-part}
\frac{1}{\sqrt{\kappa}}\del{\varphi}{\phi}-\del{\varphi}{\alpha} = \frac{1}{\mathtt{K}\E} \ .
}
By choosing $\E = 1$,\footnote{For a free relativistic particle with mass $m$ and Hamiltonian (mass-shell) constraint $g^{\mu\nu}p_{\mu}p_{\nu}/2+m^2/2=0$, where $g_{\mu\nu}$ is the spacetime metric and $c=1$, the gauge $\E = 1/m$ corresponds to choosing the particle's proper time to parametrize its trajectory. Similarly, in general relativity, the fixation of the spacetime lapse function to be equal to $1$ is related to the proper time of observers whose worldlines are orthogonal to the spatial hypersurfaces. For this reason, in the context of homogeneous cosmology or more general gravitational models, respectively, one sometimes refers to the $\E = 1$ gauge or, more generally, to $\dot{\E}=0$ (together with a suitable condition on the spacetime shift vector) as the ``proper-time'' gauge \cite{Chataignier:2023,Teitelboim:1982}. Of course, for a free relativistic particle that is massless, $\E = 1$ rather corresponds to an affine parametrization of the worldline.} we obtain
\eq{
\varphi  = \frac{\mathtt{f}(x)-\alpha}{\mathtt{K}} \ ,
}
where $\mathtt{f}$ is a dimensionless differentiable function of $\alpha+\sqrt{\kappa}\phi$. In terms of the $(s,x)$ variables, we thus have\footnote{\label{foot:units-coords}It is preferable to have coordinates on configuration space with the same units. Since $x$ is dimensionless, we can adopt $|\mathtt{K}|s$ rather than $s$ as a coordinate. Correspondingly, the configuration space is foliated by the level sets of $|\mathtt{K}|\varphi$ rather than $\varphi$. With this, one can show that the normal vector to the leaves is negativelike (and thus the leaves are positivelike, cf. Sec.~\ref{sec:fol}) if $f'(x)<1/2$ for all $x$, with $f'$ being the derivative of $f$ with respect to its argument. We can also use these dimensionless coordinates to derive the probability density given in Eq.~\eref{eg-distro-part} following Sec.~\ref{sec:probs-gen}.}
\eq{\label{eq:eg-coords-part}
\alpha(s,x) &= \mathtt{f}(x)-\mathtt{K}s \ , \ \phi(s,x) = \frac{x-\mathtt{f}(x)+\mathtt{K}s}{\sqrt{\kappa}} \ , \\
a(s,x) &= a_0\exp\left(\mathtt{f}(x)-\mathtt{K}s\right) \ ,
}
with the Gaussian distribution
\eq{\label{eq:eg-distro-part}
p_{\Psi}(x):=\frac{\sqrt{|\tilde{g}|}R^2}{\E\,\int_{-\infty}^{\infty}\D x'\,\sqrt{|\tilde{g}|}R^2/\E} = \frac{\e^{-\frac{(x-x_0)^2}{\epsilon^2}} }{\epsilon\sqrt{\pi}}\ ,
}
where $R$ is the amplitude of the wave function given in Eq.~\eref{wf-part}. In the limit $\epsilon\to0^+$, this distribution approaches a Dirac delta distribution, as it becomes very narrowly peaked at $x = x_0$. Indeed, we find the expectation values\footnote{This result holds provided taking the limit commutes with the integration, assuming that $f$ satisfies certain properties (e.g., it is continuous or smooth with compact support). To approximately obtain $\braket{f} \simeq f(x_0)$ in a similar way, if the series expansion $f(x) = \sum_{n = 0}^{\infty}f_n x^n$ holds, we see that $\braket{f} = \sum_{n = 0}^{\infty}f_nx_0^n+\Ob(\epsilon^2)$, provided we can commute the summation with the integral. If we can then truncate at lowest order in $\epsilon$, we obtain $\braket{f} \simeq f(x_0)$.}
\eq{\label{eq:dirac-delta-part}
\braket{f} &= \lim_{\epsilon\to0^+}\frac{1}{\epsilon\sqrt{\pi}}\int_{-\infty}^{\infty}\D x\,f(x)\,\e^{-\frac{(x-x_0)^2}{\epsilon^2}}\\
&= \lim_{\epsilon\to0^+}\frac{1}{\sqrt{\pi}}\int_{-\infty}^{\infty}\D x\,f(\epsilon x+x_0)\,\e^{-x^2} = f(x_0) \,.
}
In this way, the expectation values of $\alpha, a$, and $\phi$ are obtained by evaluating the expressions above at $x = x_0$. Similarly, as the expectation values of the momentum operators can be computed from the components of the gradient of the phase in Eq.~\eref{wf-part} [cf. Eq.~\eref{av-momentum-mu}], we can define the quantities
\eq{\label{eq:eg-momenta-part}
p_\alpha(s,x) &= \mathtt{K} \ ,\ p_{\phi}(s,x) = \sqrt{\kappa}\,\mathtt{K} \, ,\\
p_a(s,x) &= \frac{\mathtt{K}}{a_0}\exp\left(-\mathtt{f}(x)+\mathtt{K}s\right) \, ,
}
so that the expectation values of momenta are obtained from these functions at $x = x_0$. With this, it is straightforward to verify that these functions obey the classical equations of motion:
\eq{\label{eq:eom-part}
\del{\alpha}{s}(s,x_0) &= -p_{\alpha}(s,x_0) = \{\alpha,H\} \ , \\
\del{p_{\alpha}}{s}(s,x_0) &= 0 = \{p_{\alpha},H\} \ , \\
\del{\phi}{s}(s,x_0) &= p_{\phi}(s,x_0)/\kappa= \{\phi,H\} \ , \\
\del{p_{\phi}}{s}(s,x_0) &= 0 = \{p_{\phi},H\} \ , \\ 
\del{a}{s}(s,x_0) &= -a^2(s,x_0)p_a(s,x_0)= \{a,H\} \ , \\
\del{p_a}{s}(s,x_0) &= a(s,x_0)p_a^2(s,x_0)= \{p_a,H\} \ , \\
0&=-\frac{p_{\alpha}^2}{2}+\frac{p_{\phi}^2}{2\kappa} = H \,,
}
where $\{\cdot,\cdot\}$ is the Poisson bracket for the $(\alpha,\phi)$ variables evaluated at $x = x_0, p = \nabla S|_{x = x_0}$ (i.e., one makes this restriction after the Poisson bracket is computed). This is an elementary illustration, for the case of a very narrowly peaked state, of the Ehrenfest theorem discussed in Sec.~\ref{sec:ehrenfest}.

Incidentally, as the equations of motion~\eref{eom-part} are time-reparametrization invariant, we are free to choose another internal time, which also corresponds to changing $\varphi$ and $\E$ in Eq.~\eref{internal-time-part}. For example, if we change from the $\E=1$ parametrization to one in which $\alpha$ plays the role of a new internal ``time'' (which is dimensionless), then we see from Eqs.~\eref{eom-part} that this reparametrization is only possible if $p_\alpha = \mathtt{K}\neq0$ (which we assume). The choice $\varphi = \alpha$ in Eq.~\eref{internal-time-part} leads to $\E = -1/\mathtt{K}$, so that the restriction $\E>0$ leads to $\mathtt{K}<0$. We see from the wave function in Eq.~\eref{wf-part} that this corresponds to a restriction to a positive-frequency state of the relativistic particle, for which the Klein--Gordon squared norm is indeed positive.

\subsection{Indefinite oscillator}
Now consider the case in which $k>0$. It is then convenient to perform the change of variables\footnote{See also \cite{Greensite3}. For simplicity, we do not denote coordinates with superscripts in this Section, and we write $q_{1,2}$ instead of $q^{1,2}$.} 
\eq{\label{eq:indef-transf}
q_1 &= a\sinh(\sqrt{\kappa}\phi) \ ,\ q_2 = a\cosh(\sqrt{\kappa}\phi) \ ,
}
so that the constraint becomes proportional to that of an indefinite harmonic oscillator,\footnote{The proportionality factor is $-\kappa a^2$. This corresponds to a Weyl transformation (cf. Sec.~\ref{sec:weyl}) together with a sign inversion of the metric (and the potential), which yields $q_1$ as the negativelike direction, as we comment next.}
\eq{\label{eq:indef-wdw}
\left(\frac{\hbar^2}{2}\del{^2}{q_1^2}-\frac{m^2\omega^2q_1^2}{2}-\frac{\hbar^2}{2}\del{^2}{q_2^2}+\frac{m^2\omega^2q_2^2}{2}\right)\Psi = 0 \ ;
}
i.e., the constraint equation is equivalent to the vanishing difference of two Hamiltonians for oscillators with the same mass and frequency, where we defined
\eq{
m^2\omega^2 := \frac{k}{\kappa^2} \ ,
}
which has the correct units of mass$^2/$length$^2$ ($c = 1$). From Eq.~\eref{indef-wdw}, we note that we can take the configuration-space metric of this $k>0$ model to be ${\rm diag}(-1,1)$ in the $(q_1,q_2)$ coordinates, thus taking $q_1$ as the negativelike direction. We choose the state
\eq{\label{eq:osc-ansatz}
\Psi = \exp\left[\frac\I\hbar \frac{m\omega}{\sqrt{2}}q_1q_2-\frac{m\omega}{2\sqrt{2}\hbar}(q_1^2+q_2^2)\right]
}
as a solution to the quantum constraint.\footnote{Brotz and Kiefer considered a similar solution in \cite{Brotz} (with $\hbar = 1, m\omega = 1$): $\Psi = \exp(\I q_1q_2)$, which they examined with the Greensite--Padmanabhan assumptions, in particular $\E = 1$ and $\partial S/\partial x^i = 0$. They concluded that it was not possible to define Ehrenfest equations from this wave function, and thus that the validity of these equations is to be obtained only in restricted cases. Given the formalism presented here, we see that this wave function has $R = 1$, and therefore it cannot be normalized, not even with respect to the squared norm considered in Eq.~\eref{fol-norm}. In this way, an issue with this choice of state is that it is not normalizable, and thus it is not in the physical Hilbert space [compare with Eq.~\eref{eg-distro}].} The phase factor and amplitude are
\eq{\label{eq:phase-amp-osc}
S = \frac{m\omega}{\sqrt{2}}q_1q_2 \ ,\ R = \e^{-\frac{m\omega}{2\sqrt{2}\hbar}(q_1^2+q_2^2)} \ ,
}
and the equation for the gauge-independent variable $x$ reads [cf. Eq.~\eref{x-inv}]
\eq{
q_1\del{x}{q_2}-q_2\del{x}{q_1} = 0 \ ,
}
so we can choose
\eq{\label{eq:x-osc}
x = q_1^2+q_2^2 = a^2\cosh(2\sqrt{\kappa}\phi) \geq 0 \ .
}
As before, the internal time is a solution to [cf. Eq.~\eref{E-phase-grad}]
\eq{\label{eq:osc-E}
q_1\del{\varphi}{q_2}-q_2\del{\varphi}{q_1} = \frac{\sqrt{2}}{m\omega\E} \ .
}
For simplicity, let us choose $\varphi = -\phi$, so as to obtain
\eq{
\E = \frac{\sqrt{2\kappa}}{m\omega}\frac{(q_2^2-q_1^2)}{(q_2^2+q_1^2)} = \frac{\sqrt{2\kappa}}{m\omega\cosh(2\sqrt{\kappa}\phi)} > 0 \ .
}
In this way, the gauge choice $\varphi = -\phi$ is well defined for the solution $\Psi$. If we had chosen the gauge $\varphi = a$, we would have obtained $\E = \sqrt{2}[m\omega a\sinh(2\sqrt{\kappa}\phi)]^{-1}$, which has a discontinuity at $\phi = 0$: $\lim_{\phi\to0^{\pm}}[\sinh(2\sqrt{\kappa}\phi)]^{-1} = \pm\infty$, and thus $\varphi = a$ is not an acceptable gauge choice everywhere,\footnote{Here, we take a gauge choice to be acceptable if it leads to a continuous $\E$ that does not diverge nor vanish for finite values of the variables. See also the comments in Sec.~\ref{sec:gf}.} but only if $\phi>0$ and $a>0$ (in which case $\E >0$).

In the gauge $\varphi = -\phi$, we obtain the following expressions in the $(s,x)$ variables:\footnote{Although we choose $-\phi$ as the internal ``time,'' it does not have units of time, but rather of $(\text{mass/length})^\frac12$ with $c = 1$. For this reason, $\E$ then has units of $(\text{length/mass})^\frac32$, such that the ``time'' derivative $X$ in Eq.~\eref{phase-grad} has units of $\phi^{-1}$, as it should. If we wanted to use a true time variable defined from $\phi$, we could replace $\phi\to\sqrt{\kappa}\phi/\omega$. Then, $\E\to\omega\E/\sqrt{\kappa}$ would now have units of $\text{mass}^{-1}$ and $X$ would have units of $\text{length}^{-1} = \text{time}^{-1}$. Furthermore, as in Footnote~\ref{foot:units-coords}, we can adopt coordinates that have the same units by working with $(\sqrt{\kappa}s/\omega, \omega x)$ rather than $(s,x)$, where $s = -\phi$. With this, both coordinates have units of length $=$ time, and the configuration space can be foliated by the level sets of $-\sqrt{\kappa}\phi/\omega$. One can show that the leaves are positivelike (the normal vector is negativelike) and subsequently derive Eq.~\eref{eg-distro} with these coordinates, only reverting to the $(s,x)$ variables at the end.}
\eq{\label{eq:eg-coords}
\phi(s,x) &= -s \ ,\ a(s,x) = \sqrt{\frac{x}{\cosh(2\sqrt{\kappa}s)}} \ ,\\
q_1(s,x) &= -\sqrt{\frac{x}{\cosh(2\sqrt{\kappa}s)}}\sinh (\sqrt{\kappa}s) \ , \\
q_2(s,x) &= \sqrt{\frac{x}{\cosh(2\sqrt{\kappa}s)}}\cosh (\sqrt{\kappa}s) \ , 
}
with the distribution
\eq{\label{eq:eg-distro}
p_{\Psi}(x):=\frac{\sqrt{|\tilde{g}|}R^2}{\E\,\int_0^{\infty}\D x'\,\sqrt{|\tilde{g}|}R^2/\E} = \frac{m\omega}{\sqrt{2}\hbar}\,\e^{-\frac{m\omega x}{\sqrt{2}\hbar}} \ ,
}
where $x\geq0$ and $\D x\, p_{\Psi}(x)$ is dimensionless, as it should be. Notice that the average value of $x$ in this distribution depends on $\hbar$, $\braket{x} = \sqrt{2}\hbar/(m\omega)$, and that this distribution is not narrowly peaked at $\braket{x}$. Thus, in general, $\braket{f}\neq f(\braket{x})$. In this case, the Ehrenfest theorem establishes only the average equations of motion~\eref{Ehren-rel1} and~\eref{Ehren-rel2}, and it does not lead to the classical equations. 

As before, it is convenient to define the momenta from the gradient of the phase factor $S$. They read
\eq{\label{eq:eg-momenta}
p_\phi(s,x) &= m\omega\sqrt{\frac{\kappa}{2}} x \ ,\\
p_a(s,x) &= -m\omega\sqrt{\frac{x}{2\cosh(2\sqrt{\kappa}s)}}\sinh(2\sqrt{\kappa}s) \ ,\\
p_1(s,x) &= m\omega\sqrt{\frac{x}{2\cosh(2\sqrt{\kappa}s)}}\cosh (\sqrt{\kappa}s) \ , \\
p_2(s,x) &= -m\omega\sqrt{\frac{x}{2\cosh(2\sqrt{\kappa}s)}}\sinh (\sqrt{\kappa}s) \ .
}
One can now verify that the following relations hold
\eq{\label{eq:motion-osc}
-1 &= \del{\phi}{s} = -\frac{\E}{\kappa a^2}p_{\phi} \ ,\\
0 &= \del{p_{\phi}}{s} = \frac{\E}{\kappa a^2}\del{\mathcal{V}}{\phi} \ , \\
\del{a}{s} &= -\frac{\E}{\kappa a^2}\left(-\kappa a^2p_a\right) \ ,\\
\del{p_a}{s} &= \frac{\E}{\kappa a^2}\left[-\kappa a p_a^2+\del{\mathcal{V}}{a}\right]\ , \\ 
0 &= -\frac{\kappa a^2}{2}p_a^2+\frac{p_{\phi}^2}{2}+\mathcal{V} \,.
}
These are the equations of motion generated by the Hamiltonian $-\E(-\kappa a^2p_a^2/2+p_{\phi}^2/2+\mathcal{V})/(\kappa a^2)$, where [cf. Eq.~\eref{q-potential}]
\eq{\label{eq:q-potential-osc}
\mathcal{V} &= -\frac{ka^4}{2\kappa}-\frac{\hbar^2}{2R}\left[\del{^2R}{\phi^2}-\kappa a\del{}{a}\left(a\del{R}{a}\right)\right]\\
&= -\frac{ka^4}{4\kappa} \ ,
}
and the overall factor of $-\E/(\kappa a^2)$ is due to the fact that the original quantum constraint given in Eq.~\eref{WDW-osc} is proportional to the oscillator constraint in Eq.~\eref{indef-wdw} by a factor of $-\kappa a^2$.\footnote{The end result in Eq.~\eref{q-potential-osc} does not depend on $\hbar$ because $R$ in Eqs.~\eref{phase-amp-osc} depends on $1/\hbar$.} If we take the expectation value of the equations of motion~\eref{motion-osc} with respect to the distribution given in Eq.~\eref{eg-distro}, we find the result of Ehrenfest's theorem as expressed by Eqs.~\eref{Ehren-rel1} and~\eref{Ehren-rel2}, namely, that the equations of motion in polar form hold on average also in the relativistic case.\footnote{To be clear: we recover Eqs.~\eref{Ehren-rel1} and~\eref{Ehren-rel2} by redefining $R^2$ to be normalized. Furthermore, the quantities that here correspond to the inverse metric and potential that appear in Eqs.~\eref{Ehren-rel1} and~\eref{Ehren-rel2} are to be extracted from the Hamiltonian $-\E(-\kappa a^2p_a^2/2+p_{\phi}^2/2+\mathcal{V})/(\kappa a^2)$ defined above, which includes the overall factor of $-1/(\kappa a^2)$.} However, as the distribution is not very narrowly peaked and it depends on $1/\hbar$, the expectation values
\eq{\label{eq:av-osc}
\braket{\phi(s)} &= -s \ , \ \braket{a(s)} = \sqrt{\frac{\pi\hbar}{2\sqrt{2}m\omega\cosh(2\sqrt{\kappa}s)}} \ ,\\
\braket{p_{\phi}(s)} &= \hbar\sqrt{\kappa} \ ,\\
\braket{p_{a}(s)} &= -\frac12\sqrt{\frac{\pi\hbar m\omega}{\sqrt{2}}}\frac{\sinh(2\sqrt{\kappa}s)}{\sqrt{\cosh(2\sqrt{\kappa}s)}}
}
do not solve the classical equations of motion [in the gauge $\varphi = -\phi$ and with the classical potential equal to $-ka^4/(2\kappa)$ multiplied by the overall factor of $-\E/(\kappa a^2)$]. The same conclusion holds for the expectation values of the $(q_1,p_1;q_2,p_2)$ variables.

\section{\label{sec:conclusions}Conclusions}
The formalism of Greensite and Padmanabhan \cite{Greensite,Padmanabhan,Greensite1,Greensite2} to define time from the phase of wave function is relevant not only conceptually as a potential solution to the problem of time in relativistic quantum theories, and in quantum general relativity in particular, but also as a potential method to do practical computations in toy models of quantum gravity. Although their original approach suffered from shortcomings, as discussed by Greensite \cite{Greensite2} and by Brotz and Kiefer \cite{Brotz}, we have seen in this article how the idea of phase time can be reformulated so as to overcome the previous difficulties (cf. Sec.~\ref{sec:GP}).

Essentially, if the gradient of the phase of the wave function can be used to define a foliation of configuration space with certain assumptions (such as $N>0$ and $\E>0$), then it is possible to cast the Klein--Gordon squared norm into a positive-definite expression, which is furthermore independent of the arbitrarily chosen internal time variable. This allows us to define conserved probability distributions from the amplitude of the wave function, and thus to compute expectation values (cf. Sec.~\ref{sec:rel}). As we have seen, the probabilities can be thought of as referring to gauge-independent quantities, and the positive squared norm is gauge fixed by a version of the Faddeev--Popov procedure (cf. Sec.~\ref{sec:gf}), which is also Weyl invariant (cf. Sec.~\ref{sec:weyl}). Rather than being ``frozen'' or ``timeless'' due to the independence of the wave function on an external time, the expectation values are dynamic, and their variation is described by a version of the Ehrenfest theorem (cf. Sec.~\ref{sec:ehrenfest}).

The formalism presented here also has its limitations, and further extensions or generalizations would be required if it were to be applied to a full theory of quantum gravitation. First, the formalism is established for relativistic quantum mechanics (a ``first-quantized'' theory). Despite the fact that quantum interactions are, given our current knowledge, best described by quantum field theory (a ``second-quantized'' theory), the time-reparametrization invariant toy models, for which our phase-time approach applies, share many similarities with a canonical quantum field theory of gravitation (cf. Sec.~\ref{sec:qg}). For this reason, it is important to understand the consistency and limitations of this approach in this simplified setting. Indeed, a theory of quantum mechanics based on a Hamiltonian constraint and time-reparametrization invariance is not only useful for toy models of quantum gravity but for any situation in which a preferred external time is absent, such as the description of laboratory experiments without a preferred clock. In other words, the probabilities defined from Eq.~\eref{probs} may presumably be applied to situations in which an external observer describes the dynamics of the quantum system without reference to an external time parameter (such as the one given by the observer's watch or the laboratory clock), using instead an internal time variable.

Second, as we have discussed, narrowly peaked wave functions are needed to obtain the classical equations of motion. What is the origin of such peaked states? In ordinary quantum theory, they can arise from ``state preparations'' or ``state updates'' (in which a previous state apparently collapses to the eigenstate of some observable, such as a configuration operator). Here, we have considered an example of a peaked state (cf. Sec.~\ref{sec:eg-part}), but further examination of the notion of state updates and, indeed, of the quantum postulates is needed in order to justify the appearance of narrowly peaked states and a classical limit in time-reparametrization invariant theories. For this, a notion of decoherence relative to the physical inner product and an internal time would likely be of importance.

The generalization of the formalism to full quantum gravity will require a careful regularization of the Hamiltonian constraints, and the rigorous definition of the wave functionals that solve them. Presumably, a phase-time approach based on the phase of these wave functionals would then be possible. However, the analysis of quantum foundations becomes even more paramount for the application of such an approach to gravitation and, in particular, to cosmology (cf. Secs.~\ref{sec:qg} and~\ref{sec:eg}). For example, in the absence of an ensemble of universes, one can posit that a time-reparametrization invariant approach with probabilities \`{a} la Eq.~\eref{probs} would apply only to subsystems of the universe (each described without an external time) or, alternatively, that such probabilities do describe the quantum dynamics of the whole universe, with the distribution of possible events or measurement outcomes being related to the structure of Everett branches of the quantum state of the universe.

Moreover, it is clear that the formalism described here is also useful to pilot-wave theories, where one considers that the trajectories defined from the phase of the wave function are, in fact, real (the reader is referred to \cite{PW1,PW2} and references therein for a discussion of pilot-wave and de Broglie--Bohm theories in the context of quantum gravity). It would be interesting to apply the techniques described above to connect pilot-wave formulations of time-reparametrization invariant theories to well-defined, gauge-independent inner products and expectation values.

Third, although the expectation values computed above are dynamical, and the equations of motion are invariant under time reparametrizations, concrete calculations are predicated on an (arbitrary) choice of internal time. This choice is of course subject to the Gribov problem that generally affects gauge theories: in general, it is not possible to unambiguously fix a gauge in a global manner \cite{HT:book}. In the context of time-reparametrization invariant theories, this means that, in general, no global clock can be defined. Because of this difficulty, it is imperative to relate the phase-time formalism described here to more general gauge-theoretic constructions, such as the refined algebraic quantization programme or the BRST cohomology and their associated inner product structures (see, e.g., \cite{HT:book,Hartle}).

We thus see that there is ample opportunity to not only apply the formalism and results discussed here to different toy models of quantum gravity and cosmology, allowing the explicit calculation of dynamical expectation values and probabilistic predictions, but there are also many further developments and related lines of enquiry to be pursued. We hope to address these issues in future works.

\vspace{0.1cm}
\section*{Acknowledgments}
This work is supported by the Basque Government Grant \mbox{IT1628-22}, and by the Grant PID2021-123226NB-I00 (funded by MCIN/AEI/10.13039/501100011033 and by “ERDF A way of making Europe”). It is also partly funded by the IKUR 2030 Strategy of the Basque Government.

\end{document}